%% file: main.tex
\documentclass{aa}
\usepackage{graphicx}
\usepackage{hhline}
\usepackage{natbib}
\usepackage{txfonts}

\usepackage[usenames,dvipsnames]{color} 
\bibpunct{(}{)}{;}{a}{}{,} 
\include{newcommand} 

\newcommand\gaia{{\it Gaia}} 
\newcommand{\base}{BASE-9} 
\begin{document}
\title{
Age determination for 269 \textbf{\textit{Gaia}} DR2 Open Clusters
}
\subtitle{}
\author{
D. Bossini\inst{\ref{OAPD}}
\and
A. Vallenari\inst{\ref{OAPD}}
\and
A. Bragaglia\inst{\ref{OABO}}
\and
T. Cantat-Gaudin\inst{\ref{IEECUB}}
\and
R. Sordo\inst{\ref{OAPD}}
\and
L. Balaguer-N{\'u}{\~n}ez\inst{\ref{IEECUB}}    
\and
C. Jordi\inst{\ref{IEECUB}}
\and
A. Moitinho\inst{\ref{SIMUL}}
\and
C. Soubiran\inst{\ref{LAB}}
\and
L. Casamiquela\inst{\ref{LAB}}
\and
R. Carrera\inst{\ref{OAPD}}
\and
U. Heiter\inst{\ref{UPP}}
}
\institute{
INAF-Osservatorio Astronomico di Padova, vicolo Osservatorio 5, 35122 Padova, Italy\label{OAPD}
\and
INAF-Osservatorio di Astrofisica e Scienza dello Spazio, via Gobetti 93/3, 40129 Bologna, Italy\label{OABO}
\and
Institut de Ci\`encies del Cosmos, Universitat de Barcelona (IEEC-UB), Mart\'i Franqu\`es 1, E-08028 Barcelona, Spain\label{IEECUB}
\and
SIM, Faculdade de Ci\^encias, Universidade de Lisboa, Ed. C8, Campo Grande, P-1749-016 Lisboa, Portugal\label{SIMUL}
\and
Laboratoire d'Astrophysique de Bordeaux, Univ. Bordeaux, CNRS, B18N, all\'ee Geoffroy Saint-Hilaire, 33615 Pessac, France\label{LAB}
\and
Observational Astrophysics, Department of Physics and Astronomy, Uppsala University, Box 516, 75120 Uppsala, Sweden\label{UPP}
}
\date{Received date / Accepted date }


\abstract{\gaia\ Second Data Release provides precise astrometry and photometry for more than 1.3 billion sources. This catalog opens a new era concerning the characterization of open clusters and test stellar models, paving the way for a better understanding of the disc properties.}
{The aim of the paper is to improve the knowledge of cluster parameters, using only the unprecedented quality of the \gaia\ photometry and astrometry.}
{We make use of the membership determination based on the precise \gaia\ astrometry and photometry. We apply an automated  Bayesian tool, \base, to fit stellar isochrones on the observed \Gmag, \Bp, \Rp\  magnitudes of the high probability member stars.}
{We derive parameters such as age, distance modulus and extinction for a sample of 269 open clusters, selecting only low reddening objects {and discarding very young clusters, for which techniques other than isochrone-fitting are more suitable for estimating ages}.}
{}
\maketitle{}


\section{Introduction}
The study of the formation and evolution of Open Clusters (OC) and their stellar populations represents a backbone of research in modern astrophysics.
Indeed, they have a strong impact on our understanding of key open issues, from the star formation process, to the assembly and evolution of the Milky Way disc, and galaxies in general \citep{Friel95,Jacobson16,Cantat2016,Janes82}. 
With their ages that cover the entire lifespan of the Milky Way thin disc, OCs can be used for tracing the Galactic structure.
It is therefore essential to have precise information on a significant number of OCs, located at different Galactocentric distances together with the determination of their parameters (e.g. age, kinematics, distances, and chemistry). 
In the pre-\gaia\ era we were still far from an ideal situation: the OC census is in fact poorly known.
Currently about 3000 OCs are listed in the most recent versions of \citet[][hereafter MWSC]{Kharchenko13} and \citet[][hereafter DAML]{Dias02} catalogs. 
However, the sample is far from being complete even in the local environment \citep[within $1.6-2$ Kpc,][]{Joshi16}, where new nearby clusters are still discovered nowadays \citep[see e.g.][]{Cantat18b,Cantat18c,Castro-Ginard18}. 
At the faint end of the OC distribution, small and sparse objects and remnants of disrupted clusters can escape detection \citep{bicabonatto2011}. It is also not straightforward to distinguish true clusters from asterisms without high quality kinematic information \citep{Kos_etal18}. 
Moreover, studies on OCs may be affected by very large uncertainties on the membership, distance and metallicity and this reflects on the age determination \citep{Netopil16,Cantat18b}.

\gaia\ has opened a new era in Galactic astronomy and in cluster science, in particular, thanks to the recent second data release \citep[][hereafter GDR2]{GaiaDR2,Arenou_etal18}. GDR2 not only provides homogeneous photometric data covering the whole sky, but also unprecedented high precision kinematics and parallax information, that are fundamental to obtain accurate membership and to identify new clusters. This in turn, will allow more precise age determinations.

This paper is part of a series devoted to improve the OCs census and their parameter determination, based on \gaia\ data. \citet{Cantat18a} has derived membership probability and parameters  for  128  OCs,  by combining 2MASS photometry \citep{Skrutskie06}, \gaia\ First Data Release (DR1) TGAS parallaxes, and proper motions from either \gaia\ DR1 or UCAC4  data  \citep{Zacharias12}. \citet{Castro-Ginard18}, \citet{Cantat18b}, and \citet{Cantat18c} discovered a large number of new OCs using GDR2, and re-classified a significant number of objects that turned out to be likely asterisms and not true clusters.
\citet[][hereafter Paper~I]{Cantat18b} also updated the cluster census in the solar neighborhood, deriving memberships, mean distances and proper motions for 1229 OCs from GDR2. 
\citet{Soubiran_etal18} have made use of GDR2 to derive the kinematics of a sample of 861 OCs in the Milky Way, confirming that OCs have a similar velocity distribution to field stars in the solar neighbourhood.
This paper aims to carry out automated determination of OC parameters (age, distance, extinction) by isochrone fitting using \base\ {(\citealt{vonHippel06}, see also \citealt{Jeffery16})} on Paper~I clusters. The final catalog contains bona-fide parameters for 269 clusters. This constitutes an impressive data base to understand not only the formation and evolution of open clusters, but also the disc properties.

Section~\ref{sec:data} presents the data and the cluster selection. Section~\ref{sec:\base} summarizes the method and the priors used to derive OC parameters (i.e. \base). In section~\ref{sec:results} we present the ages and the cluster parameters obtained for our sample of OCs. Finally, section~\ref{sec:discussion} compares our results with other surveys. 

\section{Data: \gaia\ cluster selection, membership and photometry}\label{sec:data}
We make use of the cluster membership derived in Paper~I on the basis of the GDR2 photometry, proper motions and parallaxes. 
The catalog includes 1229 objects. Paper~I has indeed improved the determination of membership for their clusters, thanks to \gaia\ very precise multi-dimensional astrometric data, with proper motions precision typically of $0.05-0.3$ mas yr$^{-1}$ (for $\Gmag=14-18$) and parallaxes with precision of $\sim0.02$ mas.
We recall that to discard sources with overly large photometric uncertainties, the membership derived in Paper~I is limited to sources brighter than $\Gmag\sim$18 \citep[see][for details]{Evans18}. ite
This corresponds to the turnoff of a 3~Gyr cluster seen at 10~Kpc, assuming no interstellar extinction.
The more distant and older OCs are therefore out of our detection threshold.
We restrict our analysis to a selection of OCs, having low extinction ($A_V < 2.5$ mag) and ages older than 10~Myr  (according to MWSC and DAML catalogs). 
For younger clusters, where the unclear identification of the main-sequence turnoff (TO), contamination of Pre-MS stars, and possible age spread can compromise the isochrone-fitting method, other independent techniques are more suitable to estimate the age \citep[see, e.g.,  ][]{Bouvier18,Jeffries17a,Jeffries17b,Trevor15}. 
The final sample counts 269 OCs, located within $4.5$ Kpc.
Clearly, due to the selection criteria  we applied, our sample is far from being complete.

\begin{figure*}
    \centering
    \resizebox{\hsize}{!}{\includegraphics{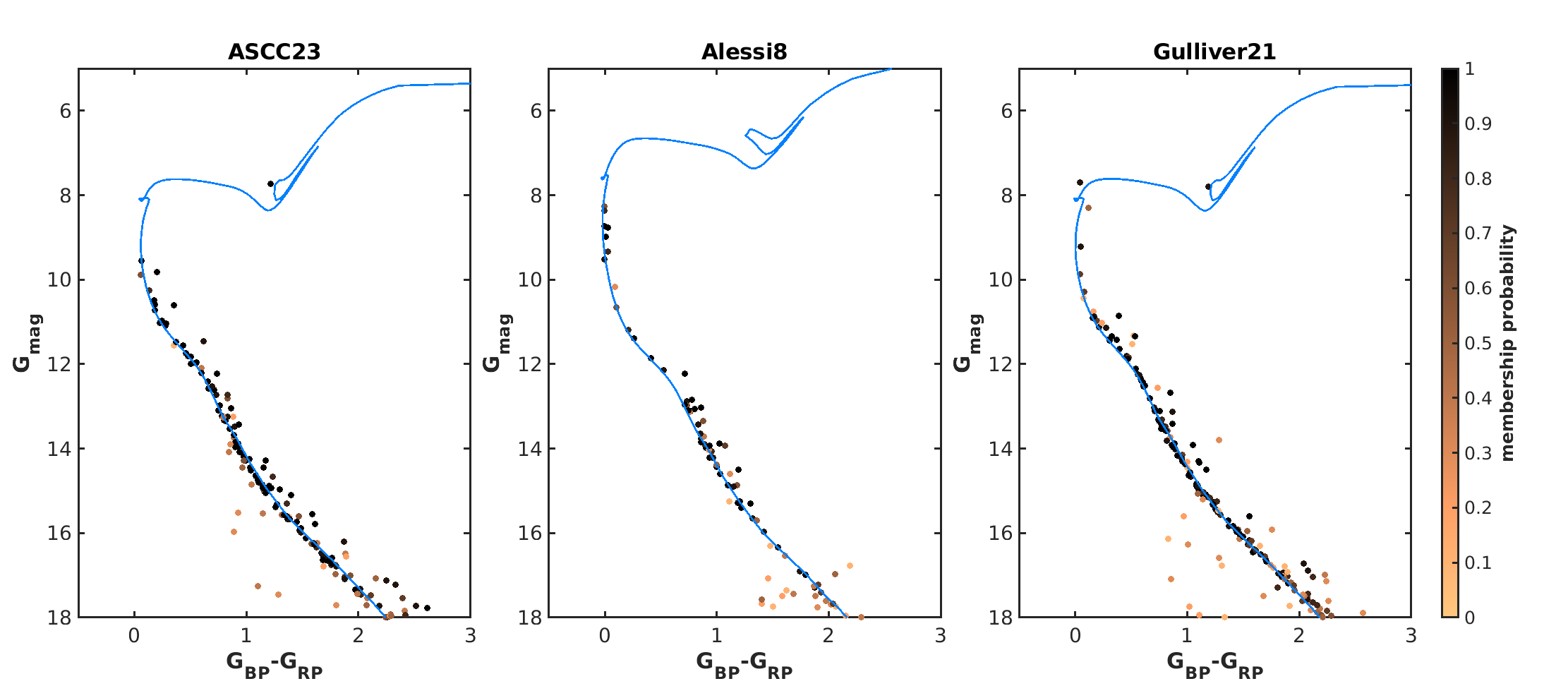}} 
    \caption{CMDs of a sample of OCs from GDR2 data, namely ASCC~23, Alessi~8, and Gulliver~21. Blue curves are the isochrones corresponding to the cluster parameters derived in this work.}
    \label{fig:cmdshow} 
\end{figure*}

In this work we make use only of the photometry from GDR2 in its three bands \Gmag, \Bp, and \Rp. This is motivated by the exceptional quality of this photometry, having a precision of the order of a few millimag \citep[see for instance][]{Babusiaux18}. Figure~\ref{fig:cmdshow} presents color-magnitude diagrams (CMD) of a few poorly studied clusters, namely ASCC~23, Alessi~8, and Gulliver~21. In all these objects, the main sequence and the equal mass binary sequence can be clearly identified.
\begin{figure*}
    \centering
    \resizebox{\hsize}{!}{\includegraphics{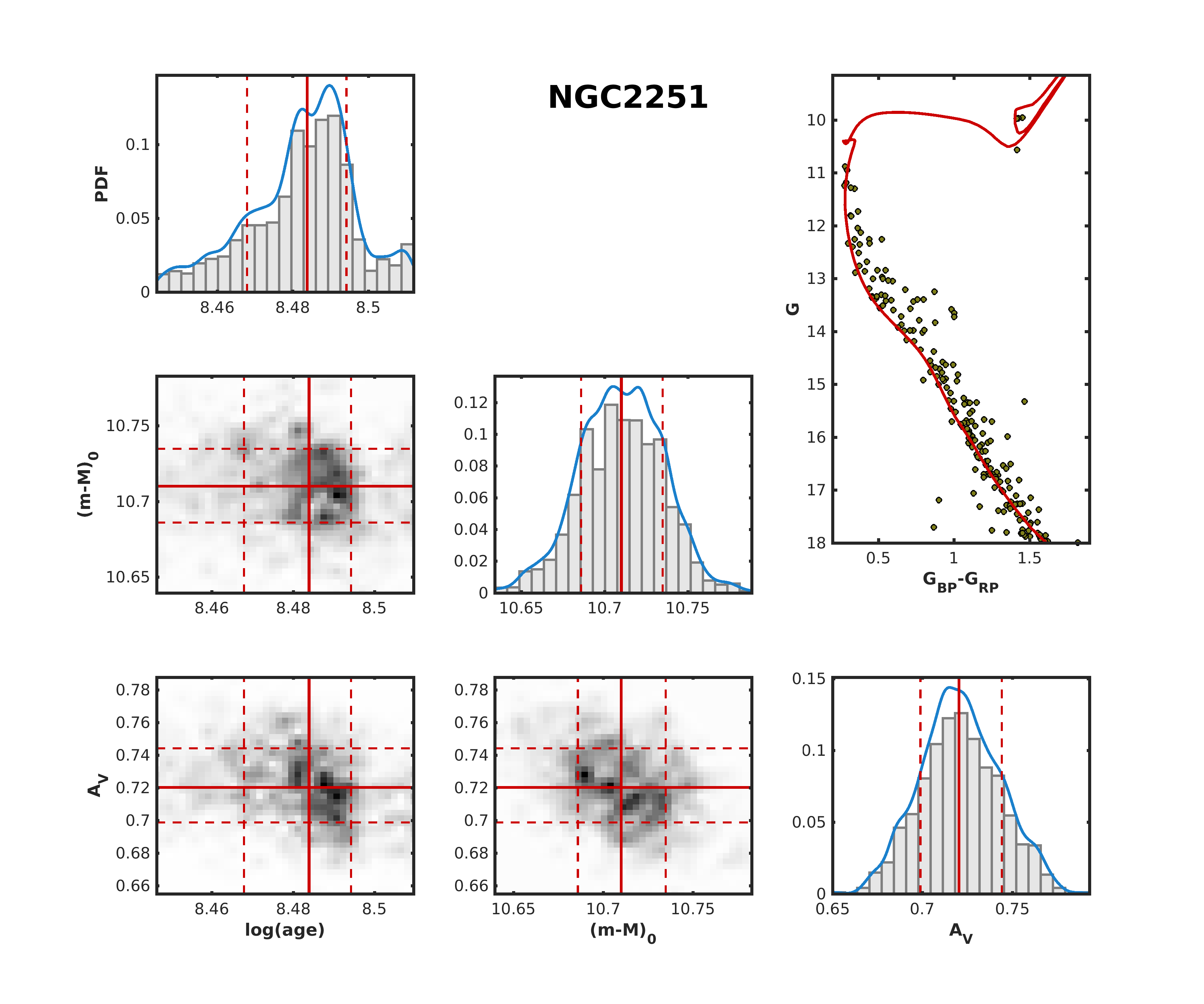}} 
    \caption{Solutions for NGC2251. On the three diagonal panels we show the probability distribution functions  of the variables with their medians (red solid line) and the 16$^\mathrm{th}$ and 84$^\mathrm{th}$ percentiles of the distribution (red dashed lines).
The panels with the maps show instead the 2D-probability for each couple of parameters, highlighting possible correlations (e.g. between distance modulus and absorption). Finally, in the top-right panel we present the CMD of the cluster with the isochrone corresponding to the median parameters.} 
    \label{fig:es1}  
\end{figure*}

\section{Method: Bayesian parameter determination} \label{sec:\base}
Handling the large, three-dimensional GDR2 datasets requires automated methods in order to characterize the OCs.  
To determine the parameters of our sample we use  an open-source software suite known as \base\ \citep{vonHippel06}, that introduces a Bayesian approach to compare observational distribution of magnitudes of stellar members of a cluster in different bands with a set of theoretical isochrones. The Bayesian method requires a likelihood function, i.e. the distribution of the data given the model parameters. 
The knowledge about the model parameters before considering the current data defines the prior distribution, while  the combined information in the data and our prior knowledge give the  posterior  distribution.
 
\base\ can adjust four parameters (age, metallicity, absorption, and distance modulus) at each iteration, using a Monte Carlo-Markov chain algorithm (MCMC). 
\base\ provides estimate of the posterior probability distribution (PDF) for a given number of iterations.  
Each iteration point is linked to the next by a ``random walk'' process described in  \citet{vonHippel06} \citet{vanDyk09}, to which we also refer for a deeper description of \base.
The introduction of priors is very useful to avoid or at least reduce local minima. Our choice  of priors is described in the section~\ref{sec:priors}. 
Visual inspection of the trace plot (parameter value against iteration number) shows that all the iteration chains reach their apparent stationary distributions within the first 1000 steps. 
This tuning period (called burn-in phase) is then discarded from the subsequent analysis. Each chain continues for an other 10,000 iterations in order to ensure statistical relevance of the results.
The clear advantage of this automated Bayesian approach for model fitting is to provide principled and reproducible estimates and uncertainties on all parameter.

The following paragraphs describe the main set-up and inputs we used in our \base\ computation.

\subsection{Stellar models and isochrones}\label{sec:model}
By default, \base\ comes with a large library of isochrones computed by different stellar-evolution groups. However none of this set of models include photometry in \gaia\ DR2 passbands \citep[][revised version]{Evans18}.
Therefore we replaced the \base-implemented PARSEC set \citep{Bressan12} with an updated version where also GDR2 passbands are available\footnote{\label{note1}\url{http://stev.oapd.inaf.it/cgi-bin/cmd}}. 
Our grid consists of isochrones in the range of $\log(age)$\footnote{logarithm of age given in years.} between 6.60 and 10.13 with a step of 0.01, and \feh\ between -2.10 and +0.50 with a step of 0.05.  
For this work we use the release PARSEC v1.2S with the bolometric corrections described by \citet{Chen14}. These authors implement the relation between the temperature T and Rosseland mean optical depth $\tau$ across the atmosphere from PHOENIX BT-Settl models as the outer boundary conditions for low temperatures. In addition  the PARSEC isochrones include a  re-calibration of the mass-radius relation for cool dwarfs as derived from eclipsing binaries. This isochrone set has been proven to reproduce not only the lower main sequence, but also all the CMD features in more than 
30 nearby OCs in GDR2 \citep[see][]{Babusiaux18}. 

\subsection{Interstellar extinction}\label{sec:extinction}
In \base\ the absorption is described using the parameter $A_V$, which is the extinction in the $V$ band. However, using different set of bands (i.e. the \gaia\ bands), it is necessary to translate $A_V$ into a proper measure of the extinction in the specific bands. 
Due to the large width of the \gaia\ bands, the coefficients $A_M/A_V$, where $M$ can be \Gmag, \Bp, and \Rp, are dependent from the stellar effective temperature \citep{Jordi10,Danielski18,Babusiaux18}. Therefore, we can expect a deviation in the shape of the reddened isochrone if a fixed relation $A_M/A_V$ is adopted.
A more sophisticated approach was introduced in \citet{Danielski18} and implemented in \citet{Babusiaux18}. In this case the extinction coefficients of the \gaia\ bands were defined as functions of the absorption $A_V$ itself and the stellar effective temperature, in the term of the color \BR:
\begin{equation}
\begin{split}
        A_M/A_V= & c_{1M} + c_{2M}\BR + c_{3M}\BR^2 + \\
                 & +c_{4M}\BR^3 + c_{5M}A_V + c_{6M}A_V^2 + \\
                 & +c_{7M}\BR A_V,
\end{split}
\label{eq:extcoefs}
\end{equation}
where $c_{1...7M}$ belong to a set of coefficients defined in \citet{Babusiaux18} for \Gmag, \Bp, and \Rp.
The terms $c_{1M}$ represents also the fixed extinction coefficients calibrated for a A0V star.
All the coefficients are listed in Table \ref{tab:coeff}.
The results presented in Tables ~\ref{Table_spectro}, ~\ref{Table_phot} and ~\ref{Table_noinfo} are given in terms of $A_{G_{TO}}$, the extinction in \Gmag\ at the turnoff of the cluster, and $A_V$, the extinction parameter to be used in equation~\ref{eq:extcoefs} to derive the dependence on color (and therefore temperature).

\begin{table}
  \caption{Summary of extinction coefficients used in this work.\label{tab:coeff}}
  \tiny{\begin{center}
    \begin{tabular}{@{}p{0.8cm}p{0.74cm}p{0.74cm}p{0.74cm}p{0.74cm}p{0.74cm}p{0.74cm}p{0.74cm}} 
    \hline        \multicolumn{8}{c}{\citet{Babusiaux18} extinction coefficients}\\
            &   $c_{1M}$  &    $c_{2M}$  &   $c_{3M}$   &   $c_{4M}$   &   $c_{5M}$   &   $c_{6M}$   &   $c_{7M}$ \\
    \hline
    $\AG/A_V$    & $0.9761$ & $-0.1704$ & $ ~~~0.0086$ & $ ~~~0.0011$ & $-0.0438$ & $0.0013$  & $0.0099$\\
    $\ABp/A_V$   & $1.1517$ & $-0.0871$ & $-0.0333$ & $ ~~~0.0173$ & $-0.0230$ & $0.0006$  & $0.0043$\\
    $\ARp/A_V$   & $0.6104$ & $-0.0170$ & $-0.0026$ & $-0.0017$ & $-0.0078$ & $0.00005$ & $0.0006$\\
    \hline
  \end{tabular}
  \end{center}}
\end{table}

\subsection{Choice of Priors}\label{sec:priors}
The Bayesian approach has the advantage that previous independent results can be incorporated through the joint prior distribution, that can be specified via independent priors on each parameter.
\base\ needs priors on the age, metallicity (\feh), $A_V$, and on the distance modulus.
To set those values we refer to literature where possible.
{  Concerning extinction, we use the values from DAML or MWSC  catalogs (prioritizing the first), where available, otherwise we set $A_V$ prior to 0.1~mag, respectively. 
$A_V$ has been marginalized within a $\sigma_{A_V}=1/3 \cdot A_V$ (or 0.033~mags if $A_V=0$).
No restriction is instead applied on the age and the variable is left free to vary inside the whole isochrone grid.}
The prior on the distance modulus is estimated through parallax inversion, which is equivalent to the equation:  
\begin{equation}
    (m-M)_0 = -5\log(\widetilde{\varpi})-5,
    \label{eq:parallax}
\end{equation}
where $(m-M)_0$ is the intrinsic distance modulus and $\widetilde{\varpi}$ is the median value of the parallax of the cluster members. 
As discussed by \citet{2018A&A...616A...9L}, such determination of distance is a very poor approximation, since systematics and correlations in the \gaia\ astrometric solution tend to overestimate the true distance, that should instead be obtained by Bayesian inferences \citep[see, e.g.,][]{2018AJ....156...58B}.
Eq.~\ref{eq:parallax} gives more consistent results for very close objects, having uncertainties on the parallax lower than 5-6\% \citep[see][]{Arenou_etal18,Babusiaux18} 
This justifies our assumption that for clusters closer than 1 Kpc and having almost no extinction, the distance modulus is assumed to be fixed to the value of equation \ref{eq:parallax}. This is the case for the clusters in common with \citet{Babusiaux18} table 2 (see also  Sect.~\ref{sec:discussion}). 
In all the other cases, the distance modulus is derived from the posterior distribution of the \base\  solutions, as recommended by  \citet{2018A&A...616A...9L}.
We must stress that \base\ looks for the observed modulus, and therefore we set the $(m-M)_V$ prior to include the contributions of both distance and extinction:
\begin{equation}
    (m-M)_V = (m-M)_0 + A_V.
    \label{eq:dist}
\end{equation}

Finally we choose to keep \feh\ fixed during the \base\ runs, in order to reduce the degeneracy within the variables.
We divide our sample in three categories having different uncertainties on the \feh\ determination: (1) clusters with high (HRS) and low resolution spectroscopy (LRS) determination of metallicity,
(2) clusters with other determination of metallicity (photometric determination, PHC), and (3) clusters with no information on metallicity (NC). 
High resolution metallicity determinations include data from \citet{Netopil16} and \gaia-ESO \citep{Spina17,Magrini17}. 
Concerning  LRS  and PHC clusters, we make use of the compilations by  \citet{Heiter14} (using spectroscopy), \citet{Paunzen10} (using photometry) as homogenized and re-calibrated by \citet{Netopil16}, who bring them on a common scale, producing the he largest homogeneous compilation of OC metallicities by far. To this group we add a few clusters whose metallicity information is taken from DAML. When this is not available we use the MWSC catalog. If  no other information is found in the literature (NC clusters), we set \feh=0.0. This is a reasonable assumption, looking at the metallicity of the OCs that are in the range $\feh \in [-0.3,+0.3]$ \citep{Netopil16}. 
The number of clusters in each group is reported in Table~\ref{tab:met_source}. 
Since using different sources for \feh\ can introduce several biases, we discuss the implication of all the above priors on the OC parameter determination in section \ref{sec:err_met}.
\begin{table}
  \caption{reference source for the metallicity priors. $N$ is the number of clusters in each group, while $\sigma_\mathrm{[Fe/H]}$ is the 90$^\mathrm{th}$ percentile of the uncertainty reported from each reference. \label{tab:met_source}}
  \tiny{\begin{center}
  \begin{tabular}{rccc}               
\hline  
 Metallicity Source & code &  $N$ & $\sigma_\mathrm{[Fe/H]}$ (dex)   \\
\hline  
 \citet{Netopil16}    & HRS & 37 & 0.09\\
  \gaia-ESO           & HRS & 4  & 0.06\\
 \citet{Netopil16}    & LRS & 3  & 0.12\\
\hline
 \citet{Netopil16}    & PHC & 25 & 0.15\\
 \citet{Dias02}       & PHC & 18 & 0.20\\
 \citet{Kharchenko13} & PHC & 3  & 0.15\\
\hline
 no \feh              & NC  & 179 & $-$\\
 \hline
  \end{tabular}
  \end{center}}
\end{table}

\subsection{Post-process analysis} \label{sec:analysis}
The probability distribution of the posteriors in \base\ are calculated during the post-process analysis.
For each run, once the chain converges, the iterations are generally distributed around a single high-probability solution. We estimate this solution through the medians for the three variables (i.e. age, extinction, and distance modulus), neglecting low probability solutions when present.
Runs having multiple (very different) solutions of comparable probability are regarded as unreliable and discarded.

It is important to notice that in principle Red Giant Branch and Red Clump stars could be used to set constrains on the metallicity. In fact, the locations of these phases on the CMD are sensitive to the change of \feh. 
However the majority of our clusters, with a few exceptions, show CMDs without these features or with only a few Red Giant stars (RG), therefore the fit is largely dominated by main sequence stars. 
As a final note, we point out that each result has been checked visually, discarding clusters with a poor isochrone fit. 


\section{Results} \label{sec:results}
The final sample of OCs span a range of $7.0$$<\log(age)<$$10.0$. Fig.~\ref{fig:es1} shows an example of the output for  NGC~2251. 
On the three diagonal panels we show the probability distribution function of the variables with their median values (red solid line), while in the top-right panel we present the CMD of the cluster where the isochrone corresponding to the median values of the solutions is overplotted. 
\begin{figure}[t]
    \centering
    \resizebox{\hsize}{!}{\includegraphics{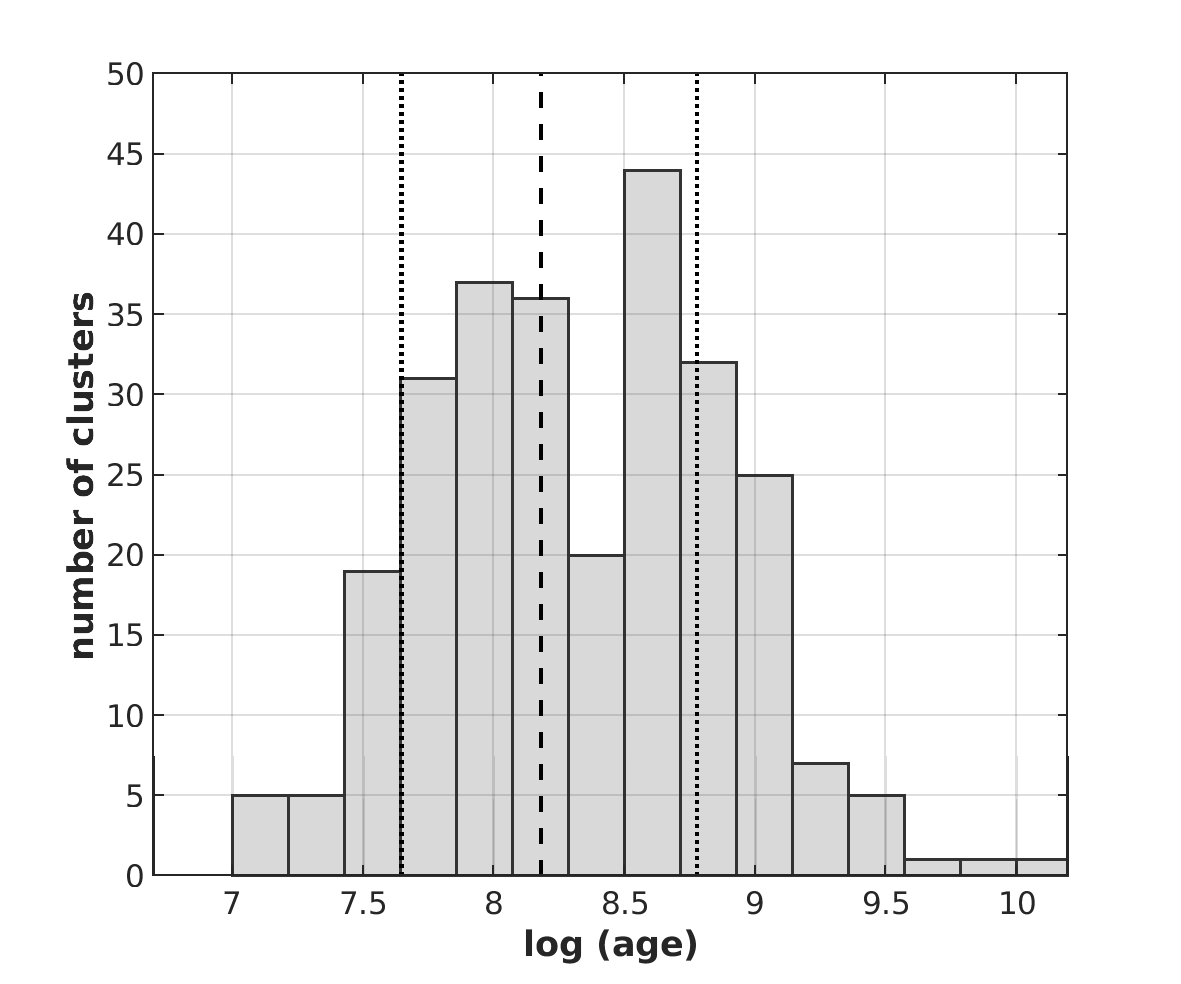}} 
    \caption{Age distribution of the studied clusters. The dashed line represents the median value of the distribution ($\log(age)=8.2$) while the dotted lines are respectively the $84^\mathrm{th}$ and $16^\mathrm{th}$ percentiles ($\log(age)_{84^\mathrm{th}}=8.8$, $\log(age)_{16^\mathrm{th}}=7.6$).} 
    \label{fig:hist_age}  
\end{figure}
The parameter determination for the three groups of clusters including the priors on \feh\ are listed in Table~\ref{Table_spectro} (44 OCs), Table~\ref{Table_phot} (46 OCs), and finally Table~\ref{Table_noinfo} (179 OCs) for  HRS+LRS, PHC, NC objects, respectively. {The values are referred to the median of each posterior distribution, while the uncertainties correspond to the 16$^\mathrm{th}$ and 84$^\mathrm{th}$  percentiles.}
Fig.~\ref{fig:hist_age} presents the age distribution of the studied clusters.

\subsection{Estimate of Uncertainties} \label{sec:errors}
In the following paragraphs we estimate the random errors on the solutions and  the systematics resulting from our assumptions on \feh.  

\subsubsection{\base\  internal uncertainties}
\begin{figure}[t]
    \centering
    \resizebox{\hsize}{!}{\includegraphics{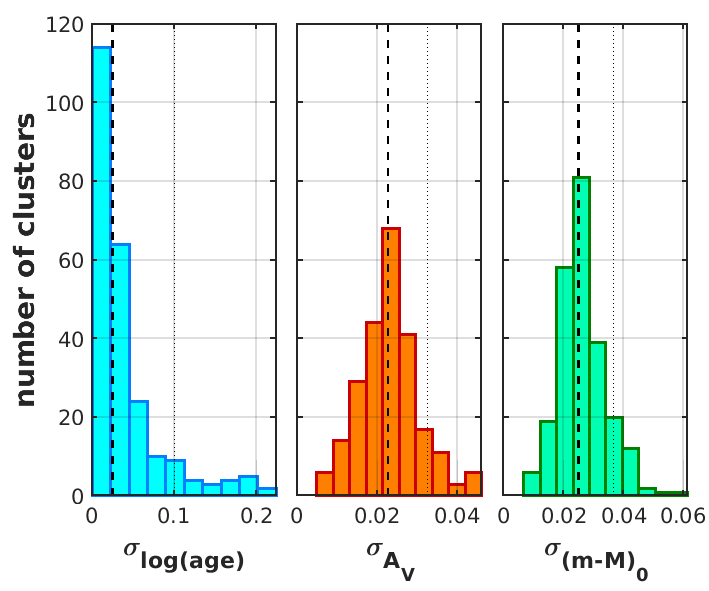}} 
    \caption{Distribution of 1$\sigma$ (estimated through percentiles) of the internal uncertainties on the age (left panel), extinction (middle panel), and distance modulus (right panel)} 
    \label{fig:INT_ERR}  
\end{figure}

{Estimation of parameter uncertainties has been done by considering the $16^\mathrm{th}$ and $84^\mathrm{th}$ percentiles (corresponding to $\pm1\sigma$) of the iterations distribution for each posterior (see Fig.~\ref{fig:es1}).} 
The distribution of the internal uncertainties for all the parameters is given in Fig.~\ref{fig:INT_ERR}.

\begin{figure}[t]
    \centering
    \resizebox{\hsize}{!}{\includegraphics{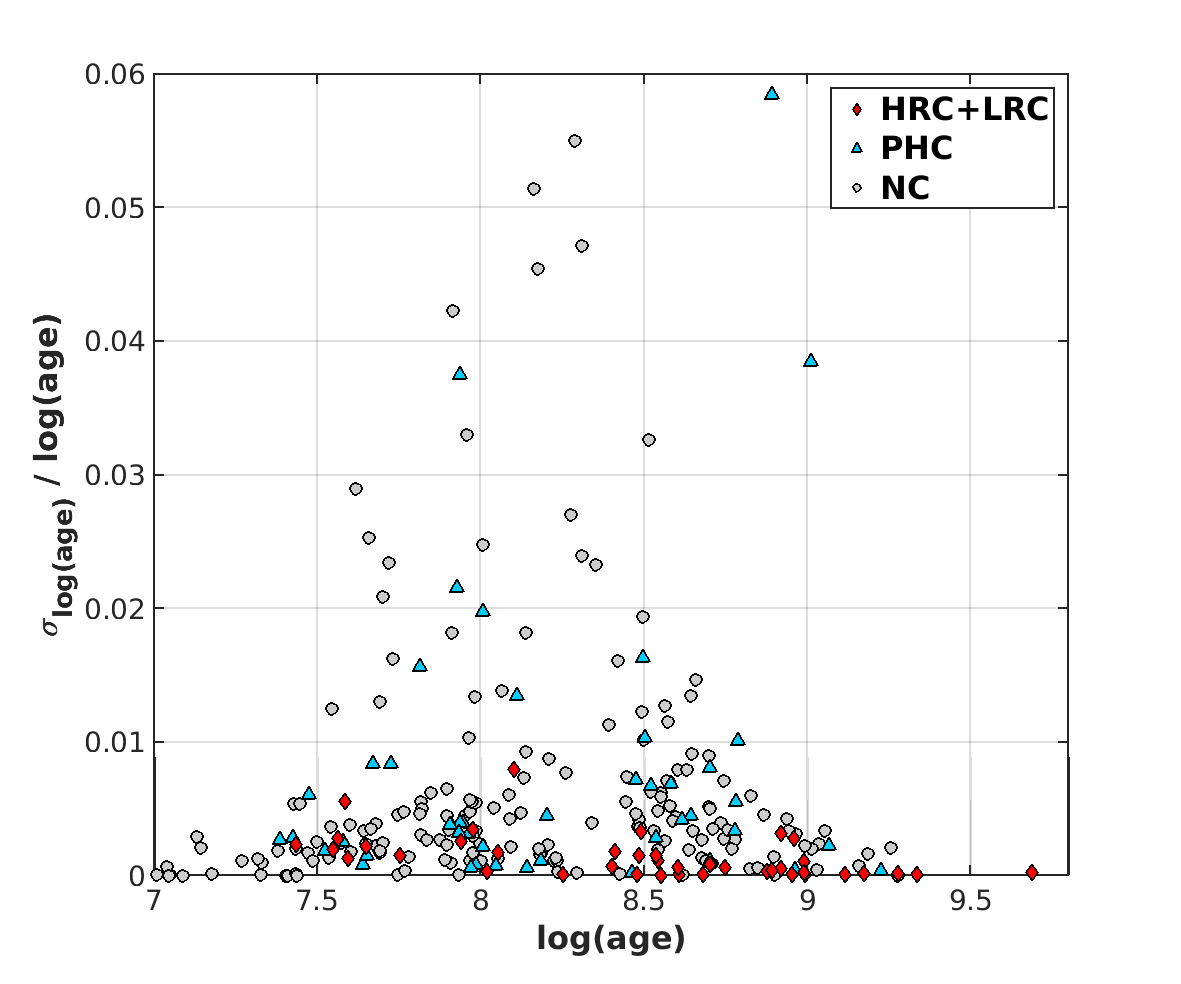}} 
    \caption{\base\  relative error on $\log(age)$ against $\log(age)$. } 
    \label{ageVserr}  
\end{figure}

We find that 90\% of the clusters  have sigmas smaller than, respectively, $\sigma_{\log{(age)}}=0.10$, $\sigma_{A_V}=0.033$, and $\sigma_{(m-M)_0}=0.037$, while their medians are $\widetilde{\sigma}_{\log{(age)}}=0.024$, $\widetilde{\sigma}_{A_V}=0.023$, and $\widetilde{\sigma}_{(m-M)_0}=0.025$.
While the extinction and the distance modulus determination are well confined, the distribution of the uncertainties on the $\log(age)$ presents a tail of about 30 OCs having $0.1< \sigma_{\log(age)} < 0.25$. 
Typical examples of this category of objects are Gulliver~20, IC~2157, and Ruprecht~29. These clusters are characterized by having no information on \feh\ (i.e we assume \feh=0.0); high extinction ($A_V > 1.0$) and a low number of members. 
For these reasons, their fits are not well constrained, and the solutions present a high degree of degeneracy between the extinction and the distance modulus. 
Fig.~\ref{ageVserr}  and Table~\ref{tab:age_error} shows the distribution of the relative error on $\log(age)$.  
HRS and LRS clusters have smaller internal uncertainties, while clusters belonging to the PHC group present larger errors. 
We detect no trend of $\sigma_{\log(age)}$ as a function of the $\log(age)$.  
\begin{table}
\caption{\label{tab:age_error} Internal uncertainties on the \base\  log(age) determination for the different groups of OCs, namely those having \feh\ from spectroscopy (both HRS and LRS), photometry, or no information respectively}
  \tiny{\begin{center}
  \begin{tabular}{c|llll}               
  \hline     $\log(age)$ & \multicolumn{4}{c}{median $\sigma/\log(age)$ \,\,\,({\it number of clusters})}   \\
             interval    & HRS+LRS           & PHC               & NC                 & all                 \\
 \hline      $7.0-7.7$   & 0.0023 ({\it ~6}) & 0.0027 ({\it ~8}) & 0.0019 ({\it ~41}) & 0.0021 ({\it ~55})  \\
             $7.7-8.5$   & 0.0017 ({\it 12}) & 0.0039 ({\it 22}) & 0.0047 ({\it ~83}) & 0.0040 ({\it 117})  \\
             $8.5-10.0$  & 0.0003 ({\it 26}) & 0.0051 ({\it 16}) & 0.0034 ({\it ~55}) & 0.0027 ({\it ~93})  \\
\hline                   &                   &                   &                    &                     \\
[-1em]\hline $7.0-10.0$  & 0.0007 ({\it 44}) & 0.0039 ({\it 46}) & 0.0034 ({\it 179}) & 0.0029 ({\it 269})  \\
\hline
 \end{tabular}
 \end{center}}
\end{table}


\subsubsection{Impact of fixed-metallicity prior}\label{sec:err_met}
As discussed in Sect.\ref{sec:priors}, we use a fixed metallicity in our \base\ calculations, and this can have an impact on results.  The aim of this section is to estimate the degree of degeneracy between the parameter determination and  \feh.

\begin{figure}[t]
    \centering
    \resizebox{\hsize}{!}{\includegraphics{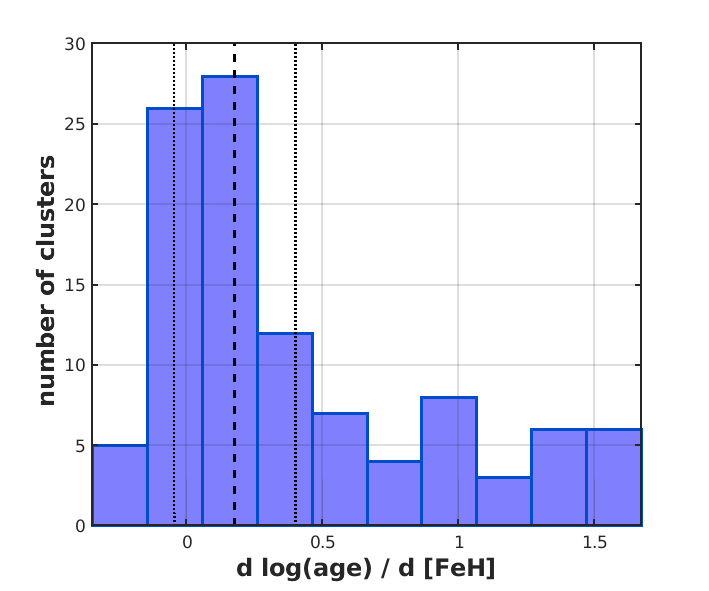}} 
    \caption{Distribution of $\D\log(age)/\D\feh$, imposing several $\feh$ values as prior for a sample of about 100 objects. Dashed line correspond to the median, while dot lines are the median $\pm$ the MAD. }
    \label{fig:ERR_LOGAGE_M}  
\end{figure}

We select from our catalog a sample of $\sim$100 clusters spanning the whole age range we consider, and we run \base\ on them using three different priors on the metallicity, $\feh=-0.3,-0.1,+0.1$. In this test, the metallicity is let free to vary within a $\sigma_{\tiny\feh}=0.05$.
Using these three runs, we calculate for the three solutions of each cluster the regression line on the plane $\log(age)-\feh$. The slope $\D\log(age)/\D\feh$ gives the predicted variation of the $\log(age)$ within 1 dex in metallicity for that specific cluster. 
Considering the overall distribution, we find a median slope of 0.18 with a median absolute deviation (MAD) of 0.22 (see Fig.~\ref{fig:ERR_LOGAGE_M}). 
Clearly the systematics we introduce on the OC parameter determination are different depending on the uncertainties on the \feh\ priors (see Table \ref{tab:met_source}). The effect can be negligible in the case of objects having \feh\ determination from high resolution spectroscopy. Assuming as typical  the sigma of 0.06 dex on \feh\ determination as derived from high resolution spectroscopy in the \gaia\ ESO public survey \citep[see for instance][]{Jacobson16}, we obtain  $$\Delta\log(age)=\dfrac{\D\log(age)}{\D\feh}\cdot\Delta\feh= \pm 0.01.$$ 
As we mentioned, for the NC group we assume \feh=0.0.  Looking at the distribution of the metallicity of Galactic clusters, we  expect  that all objects are inside a $\Delta\feh= \pm 0.3$. In this case we estimate an effect on $\log(age)$ of $\pm 0.05$, which, translated in linear age, corresponds to  about $13\%$.
Clusters having \feh\ from photometry or low resolution spectroscopy can be regarded as having intermediate uncertainties.
In the case of PHC objects, we find a median value of  $\sigma_{[Fe/H]}=0.15-0.20$, resulting in $\Delta\log(age) =0.03-0.04$, while for the LRS sample we derive $\sigma_{[Fe/H]}=0.12$, corresponding to $\Delta\log(age) =0.02$.

The apparent distance modulus variations at changing \feh\ are quite small, with a median value $\Delta{(m-M)=0.012 \pm 0.007}$ for the extreme case when we assume \feh=0.0. However, the extinction and the absolute distance modulus solutions are more affected by the assumption on \feh, with a clear degeneracy. We find a median value of $\Delta A_V=0.19 \pm 0.02$ and $\Delta(m-M)_0= 0.2 \pm 0.03$ for the extinction and the absolute distance modulus respectively. 

\section{Discussion}\label{sec:discussion}

\subsection{Comparison with benchmark clusters}
We compare our results with a set of well studied clusters having high quality determination of the parameters.
\subsubsection{Nearby OCs}\label{sec:CC}
\begin{figure}[t]
    \centering
    \resizebox{\hsize}{!}{\includegraphics{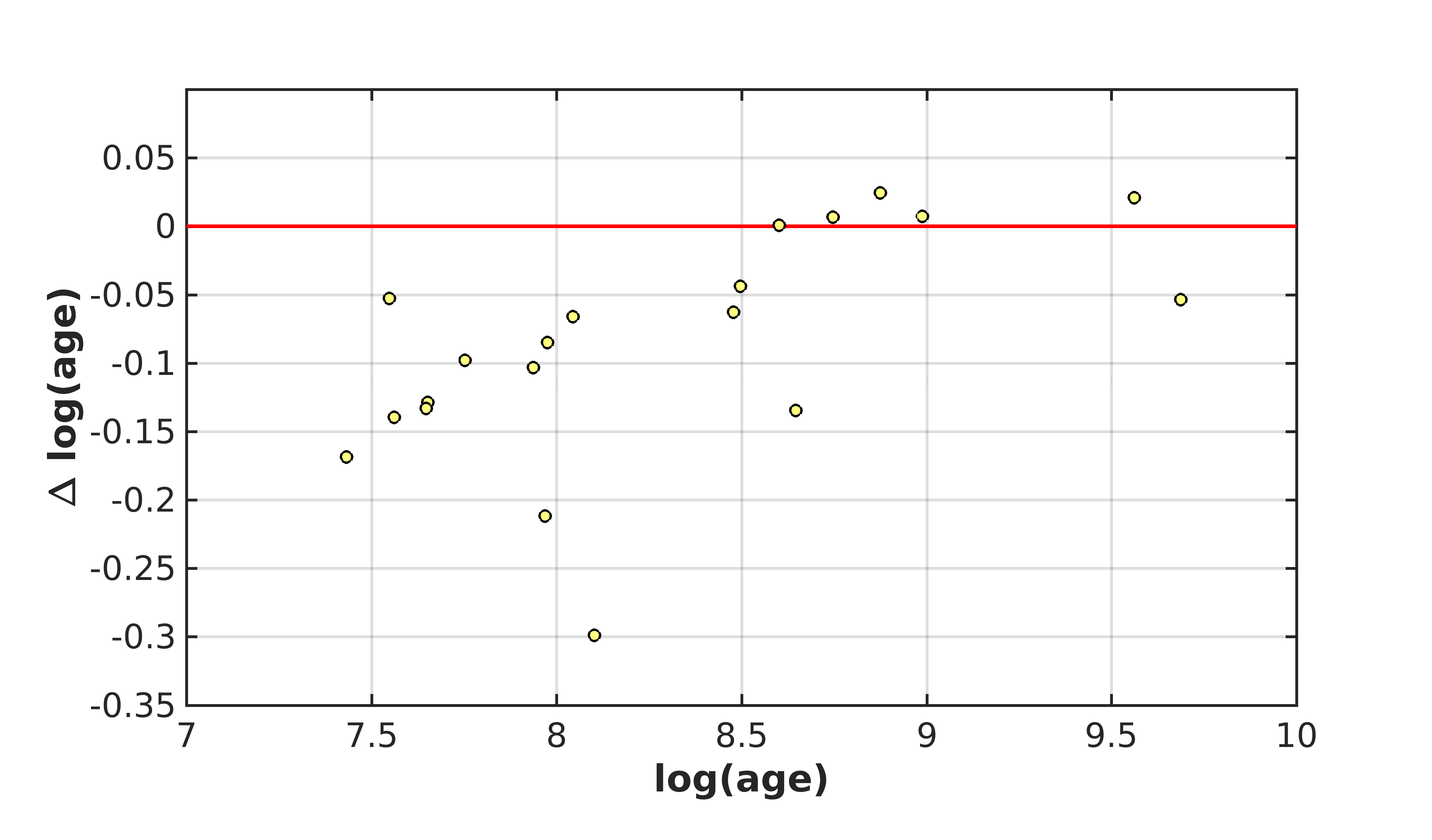}} 
    \caption{Comparison between the age derived in this work against \citet{Babusiaux18} for 20 clusters in common. $\Delta\log(age)= \log(age)_\mathrm{this work}-\log(age)_\mathrm{ref}$ versus $\log(age)_\mathrm{this work} $.}
    \label{fig:OC_babu}
\end{figure}

\begin{figure*}
\begin{minipage}{0.24\textwidth}
\centering
\resizebox{\hsize}{!}{\includegraphics{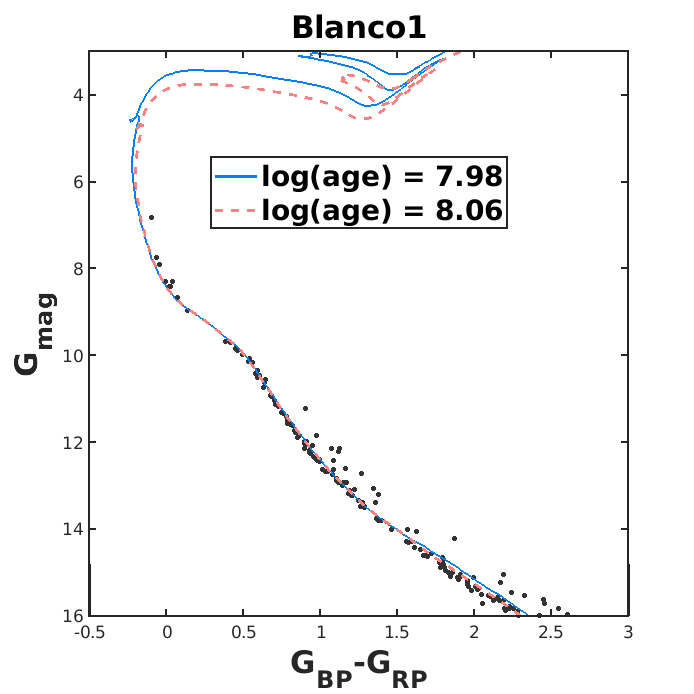}}
\resizebox{\hsize}{!}{\includegraphics{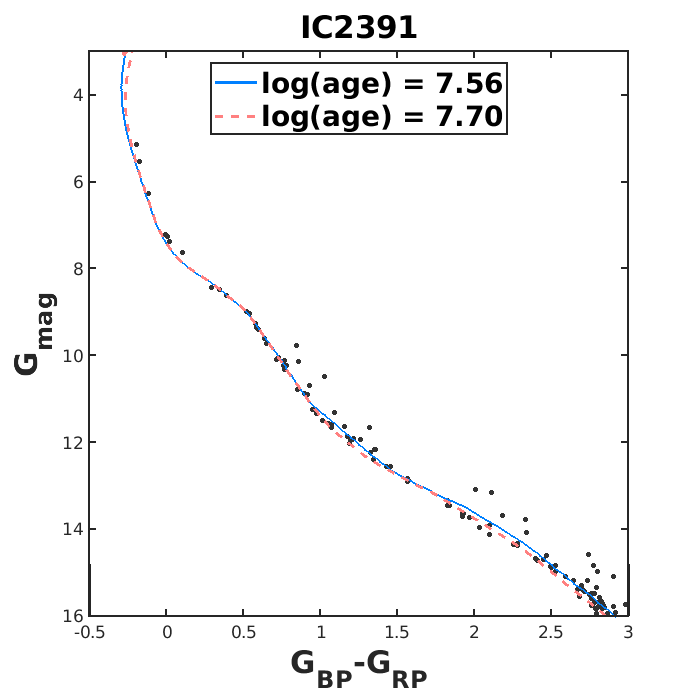}}
\resizebox{\hsize}{!}{\includegraphics{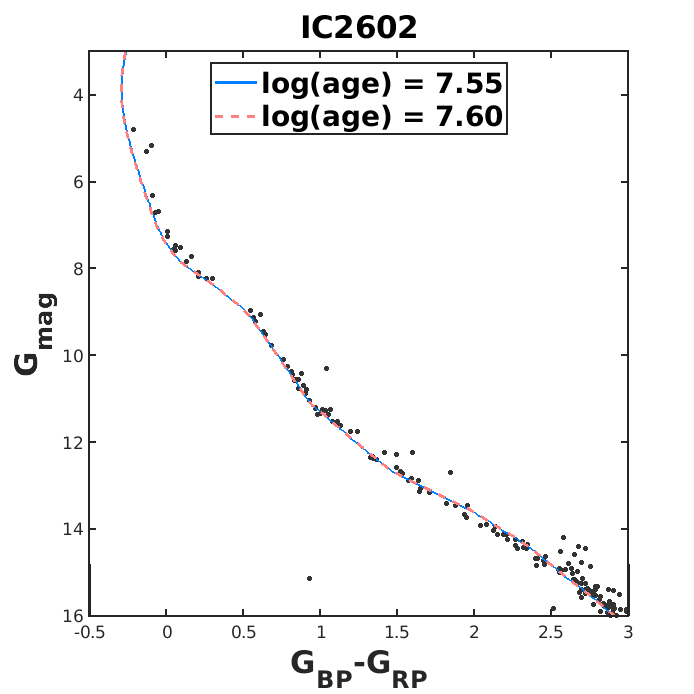}}
\resizebox{\hsize}{!}{\includegraphics{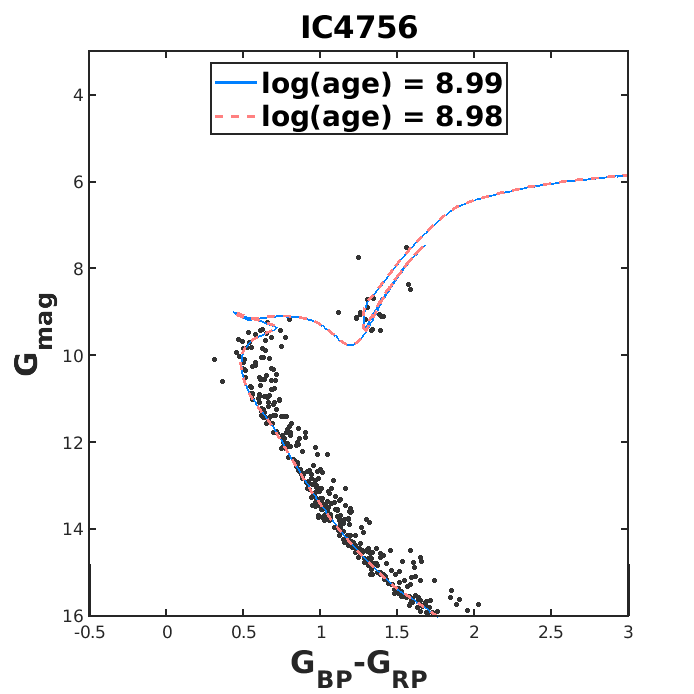}}
\resizebox{\hsize}{!}{\includegraphics{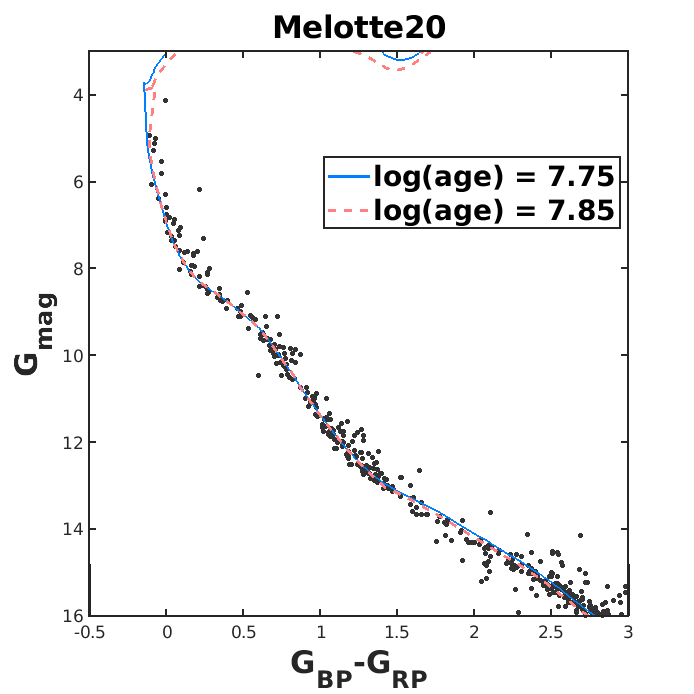}}
\end{minipage}
\begin{minipage}{0.24\textwidth}
\centering  
\resizebox{\hsize}{!}{\includegraphics{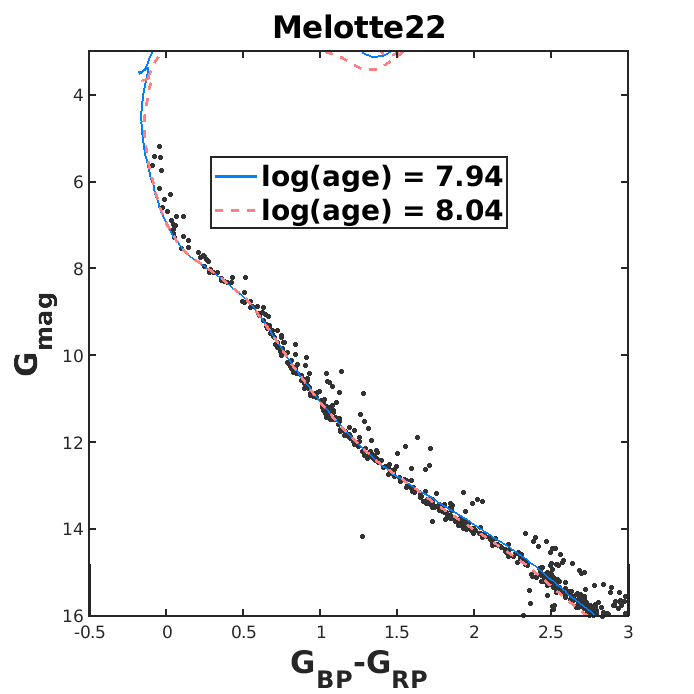}}
\resizebox{\hsize}{!}{\includegraphics{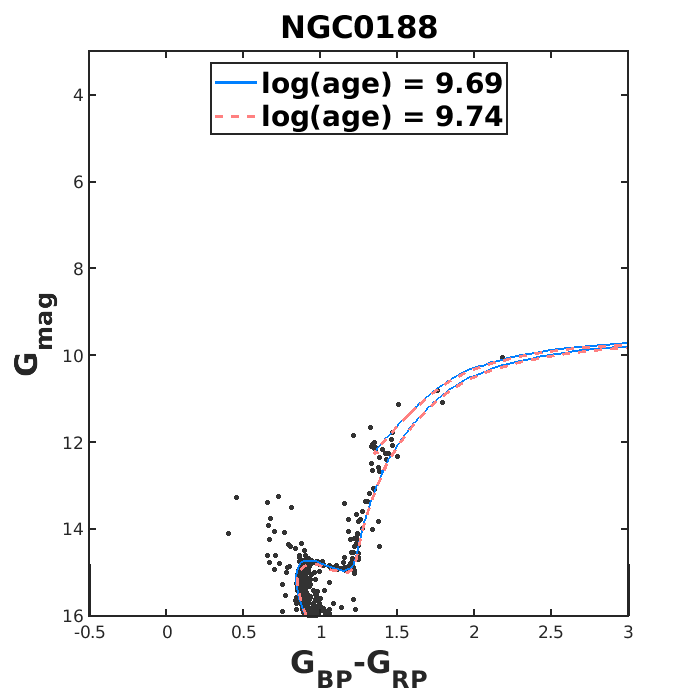}}
\resizebox{\hsize}{!}{\includegraphics{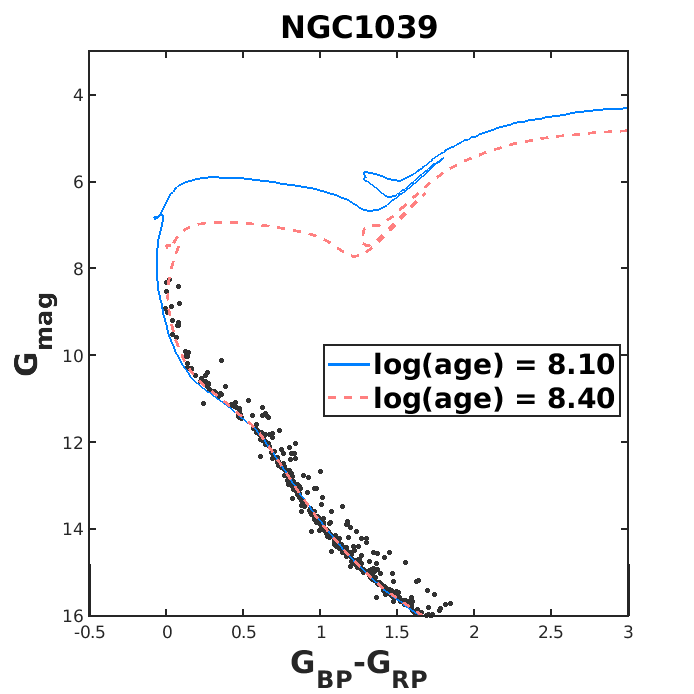}}
\resizebox{\hsize}{!}{\includegraphics{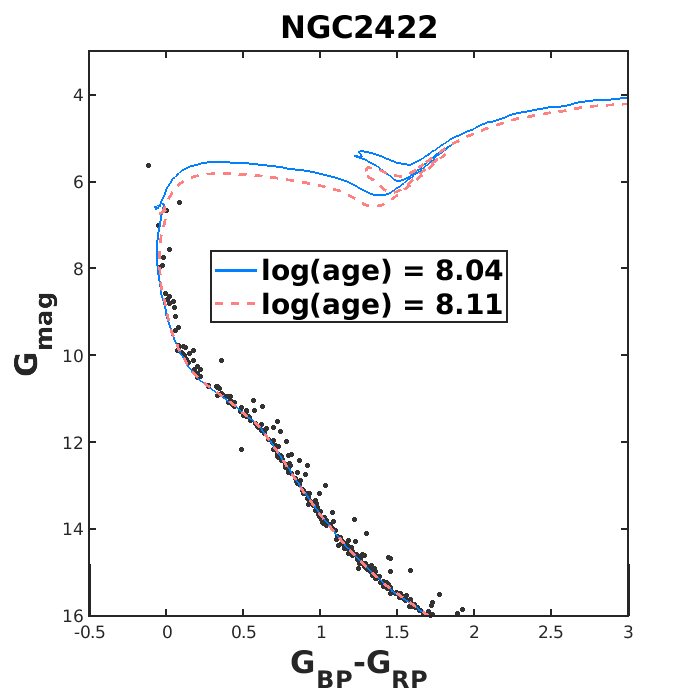}}
\resizebox{\hsize}{!}{\includegraphics{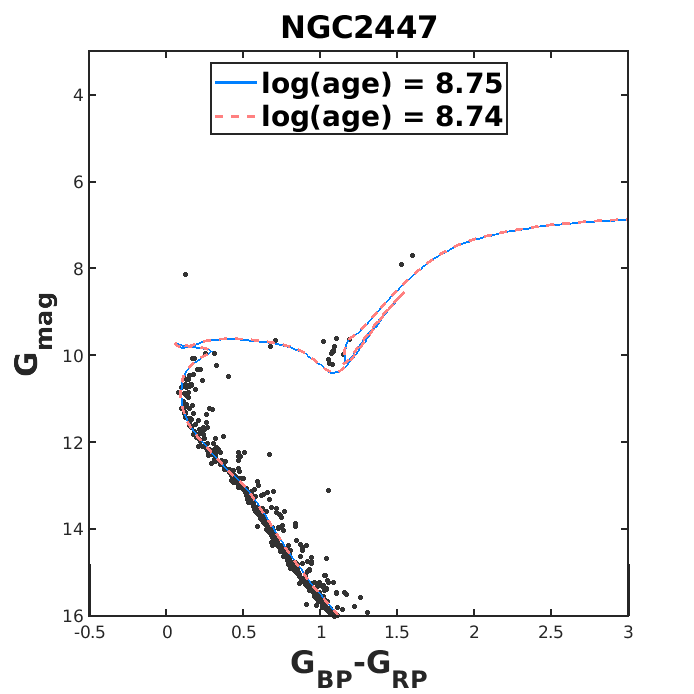}}
\end{minipage}
\begin{minipage}{0.24\textwidth}
\centering  
\resizebox{\hsize}{!}{\includegraphics{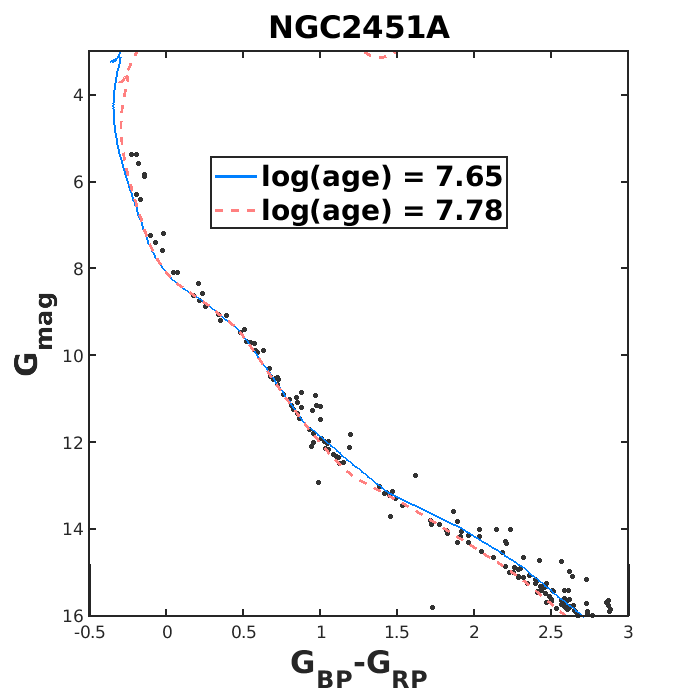}}
\resizebox{\hsize}{!}{\includegraphics{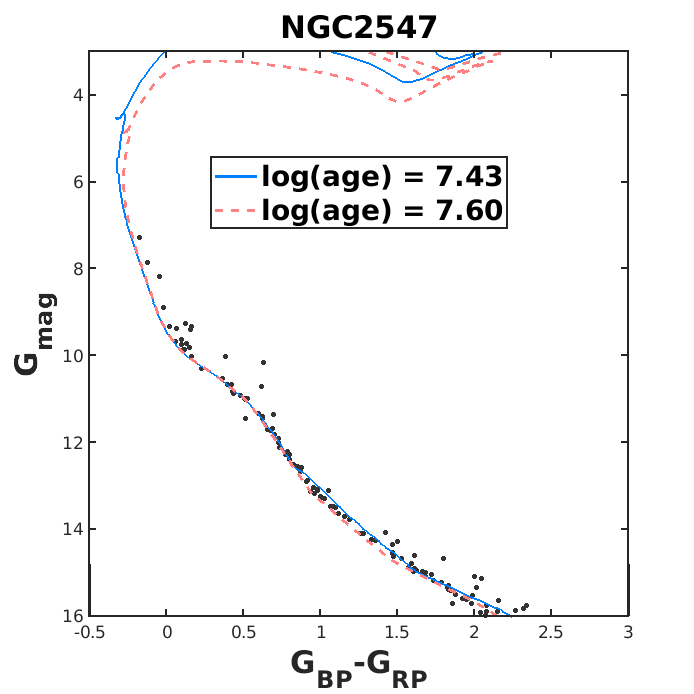}}
\resizebox{\hsize}{!}{\includegraphics{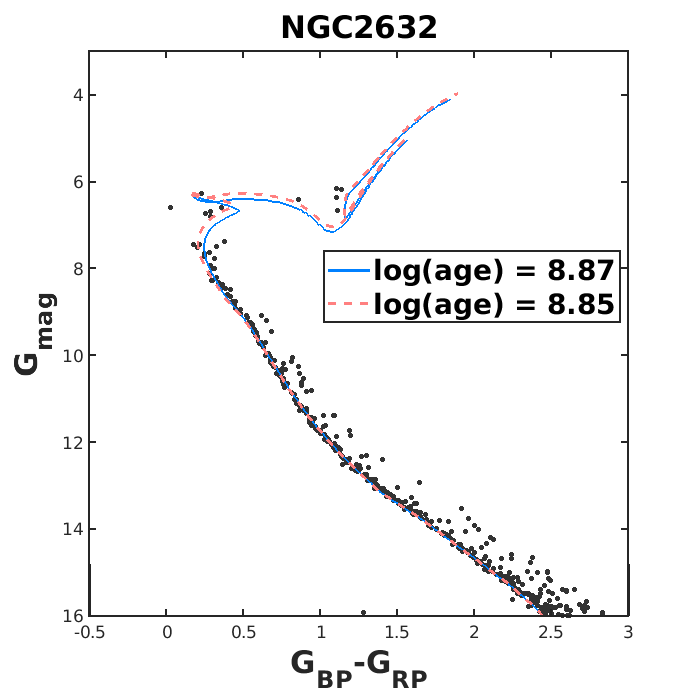}}
\resizebox{\hsize}{!}{\includegraphics{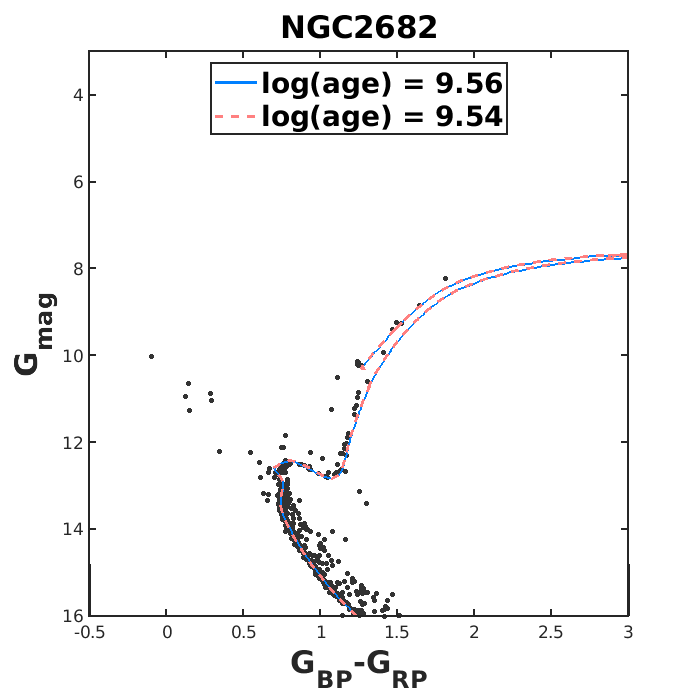}}
\resizebox{\hsize}{!}{\includegraphics{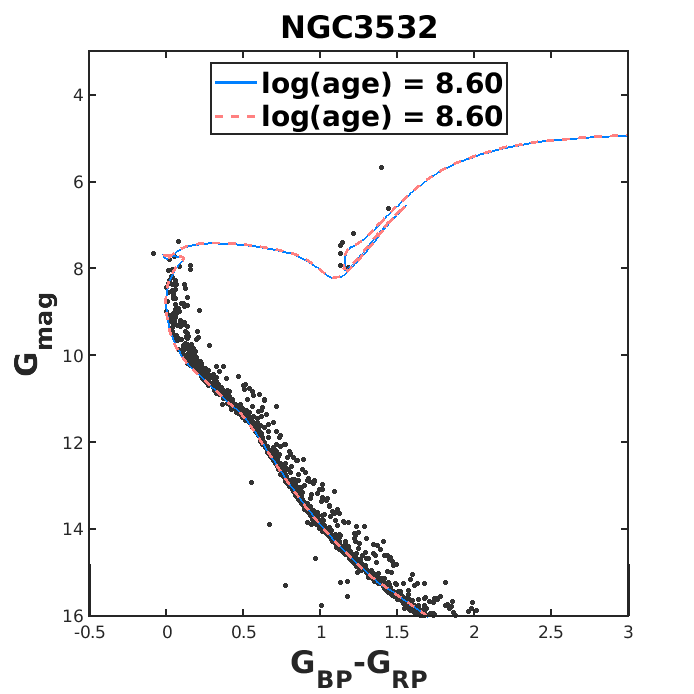}}
\end{minipage}
\begin{minipage}{0.24\textwidth}
\centering  
\resizebox{\hsize}{!}{\includegraphics{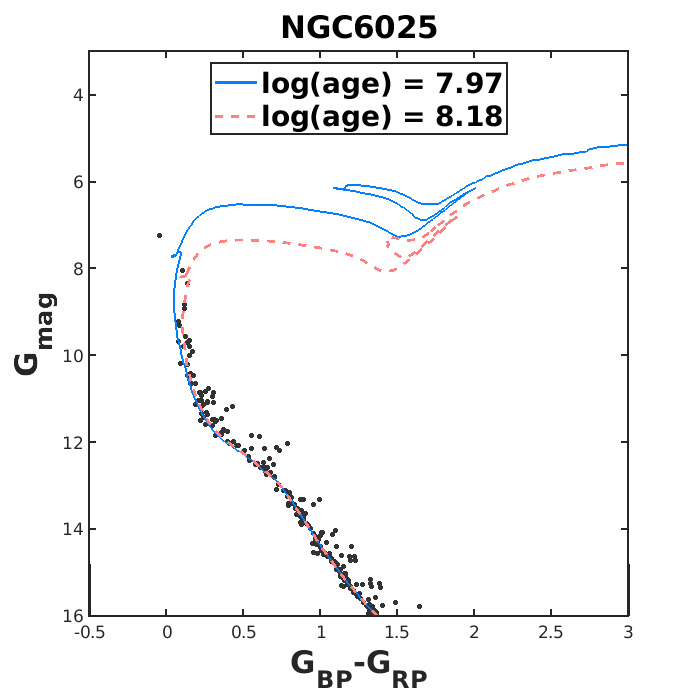}}
\resizebox{\hsize}{!}{\includegraphics{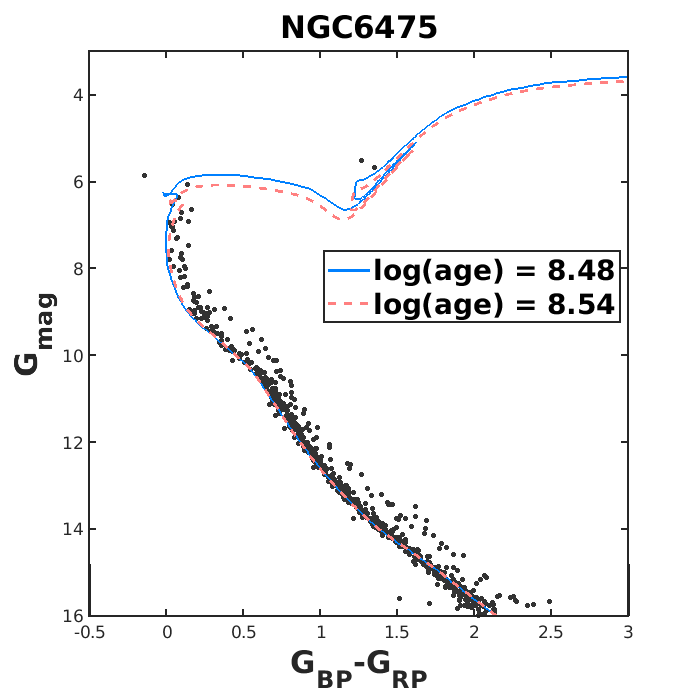}}
\resizebox{\hsize}{!}{\includegraphics{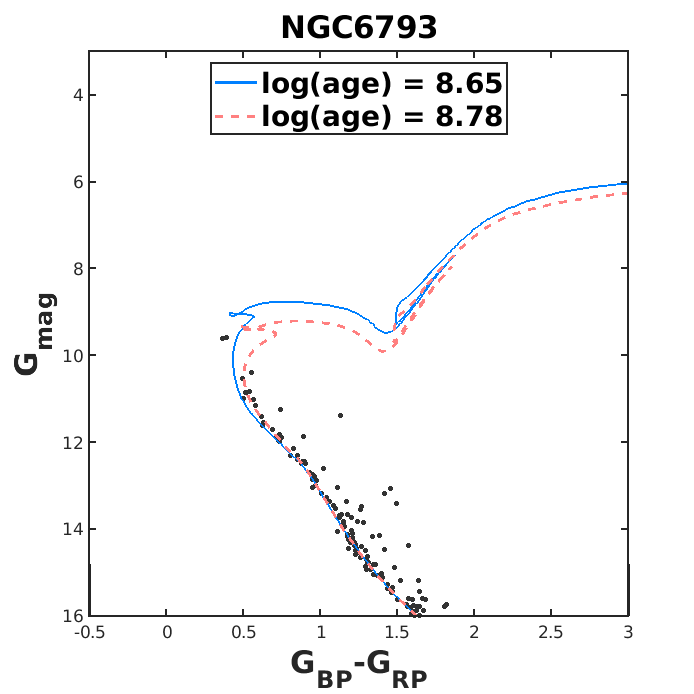}}
\resizebox{\hsize}{!}{\includegraphics{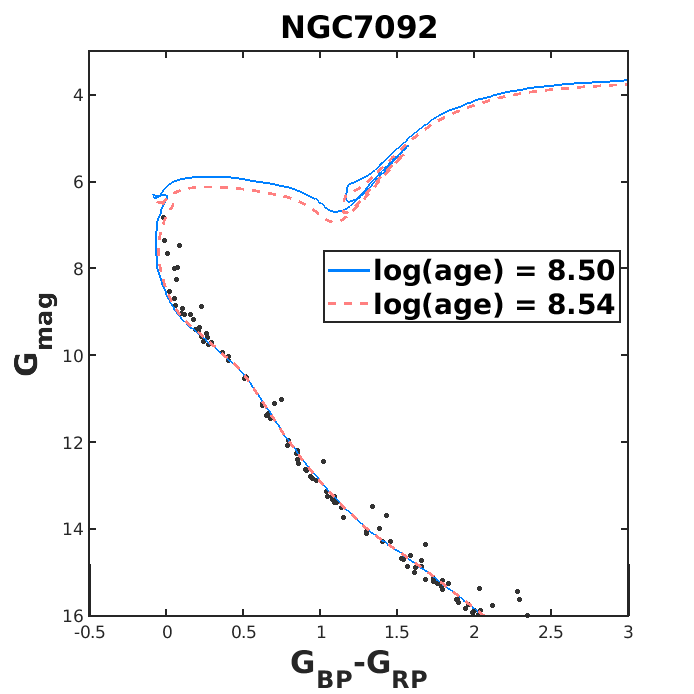}}
\resizebox{\hsize}{!}{\includegraphics{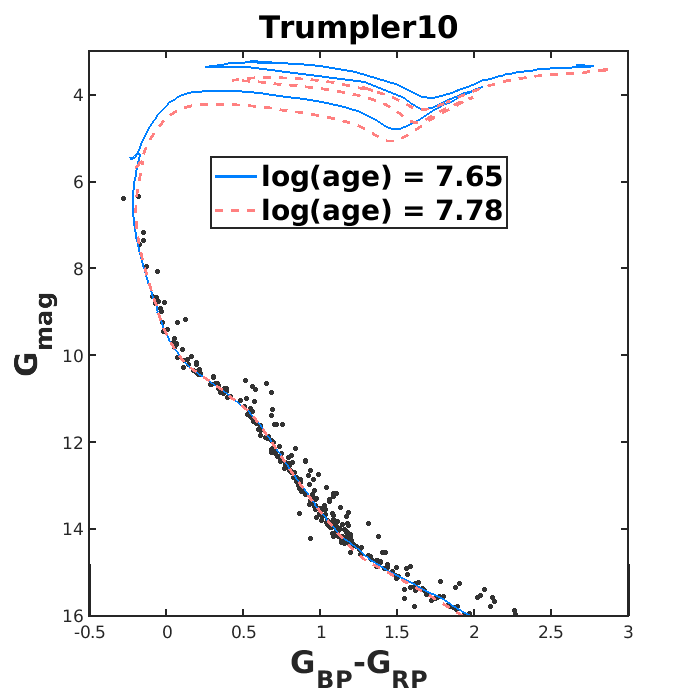}}
\end{minipage}
\caption{CMDs of the clusters in common with \citet{Babusiaux18}. Blue and red lines represent, respectively, the isochrones with the age proposed in this work and by \citet{Babusiaux18}.}
\label{fig:CC} 
\end{figure*}

In our sample we have 20 nearby clusters already studied by the \citet{Babusiaux18} using GDR2 data. 
In Fig.~\ref{fig:OC_babu}  we compare the age determination in both papers.
Both determinations agree within a few percent, showing however a small systematic underestimate for the younger objects. This deviation is mainly due to the fact that the majority of the clusters are inconspicuous (see Fig.~\ref{fig:CC}).
In some cases differences between the two ages can be ascribed to the membership determination. One example can be NGC~6793 (see Fig.~\ref{fig:CC}). In this very poorly studied cluster, the bright star at $\Gmag\sim9$ has a high probability membership from Paper~I, while in the \citet{Babusiaux18} it is not considered as a member: this changes the position of the MSTO and, therefore, the age from $\log(age)=8.78$ to $8.65$. 
We find a similar trend also  when comparing with MWSC and DAML (see Section~\ref{sec:lett}).
\subsubsection{Comparison with asterosesimic data}
Our sample contains also three OCs studied by {\it Kepler} \citep{Borucki_etal10}, i.e. NGC~6791, NGC~6811, and NGC~6819.
In many of their red-giant stars, solar-like oscillations have been detected, providing global seismic parameters such as the large separation \Dnu\ and the frequency of maximum oscillation power \numax.  These quantities, combined with the effective temperature, can be used to derive stellar masses through the so-call scaling relations (see, e.g., \citealt{Kallinger10} and \citealt{Mosser10}).

In turn, the mass can be used to provide an indirect validation of our age determination.
We compare previous estimations of RG masses for these clusters with the range of values corresponding to the same evolutionary phases along our isochrones.

\paragraph{{\bf NGC~6791}}
Seismic determination of the average RGB mass for NGC~6791 gives a value of $M=1.22\pm0.01$ \Msun\ \citep[][considering stars up to the RC luminosity]{Miglio12}. 
However, it was demonstrate that scaling relations tend to overestimate the value of the mass of RGB stars (\citealt{White11}, \citealt{Brogaard18} and reference therein), therefore an additional calibration  is required \citep{Rodrigues17}. This introduces a systematic on the mass of $\pm0.10$ \Msun.
The estimation of the mass from RGB eclipsing binaries is $M=1.15\pm0.02$, in agreement with the previous seismic determination \citep{Brogaard12}. 

Both the measures are perfectly compatible with the mass of $M=1.13\pm0.01$ \Msun, as derived averaging the masses from the bottom of the RGB up to the RC luminosity for an isochrone of the age of $\log(age)= 9.927\pm0.002$, corresponding to our solution.

\paragraph{{\bf NGC~6811}}
\citet{Sandquist16} determined the masses for 6 stars, 5 of them belonging to the red-clump phase, plus 1 RC candidate. 
The average value is $M=2.24\pm0.07$ \Msun, which is compatible with our average mass determination of $M=2.31\pm0.08$ \Msun\ for the red-clump phase in an isochrone of $\log(age)=8.94$. 

\paragraph{{\bf NGC~6819}}
\citet{Handberg17} derived seismic parameters for 54 RG stars in NGC~6819.
Within the sample, they were able to distinguish between RGB and RC stars. They also   identified non-member stars (3), stars classified as overmassive (6), uncertain cases (5), and 1 Li-rich RC.
In a subsequent work,  \citet{Rodrigues17} estimated individual masses and ages for 52 RC stars. 
They compared observational data, including seismic constrains from \citet{Handberg17}, with a grid of models through a Bayesian method \citep[PARAM,][]{daSilva06}.
Using only single RGBs they found an average mass of $M_\mathrm{RGB} = 1.61\pm0.04$ \Msun. 
Our \base\ solution for NGC~6819 corresponds to an isochrone of  $\log(age)=9.30$, that gives an average mass for the RGB of $M = 1.675\pm0.005$. This values shows only a partial compatibility with \citet{Rodrigues17} determination, lying within $1.5 \sigma$, since they find an age of $\log(age)=9.35\pm0.03$ using a different set of stellar models by  \citet{Bossini15}.  

\subsection{Comparison with MWSC and DAML catalogs}  \label{sec:lett}

\begin{figure*}
    \centering
    \resizebox{\hsize}{!}{\includegraphics{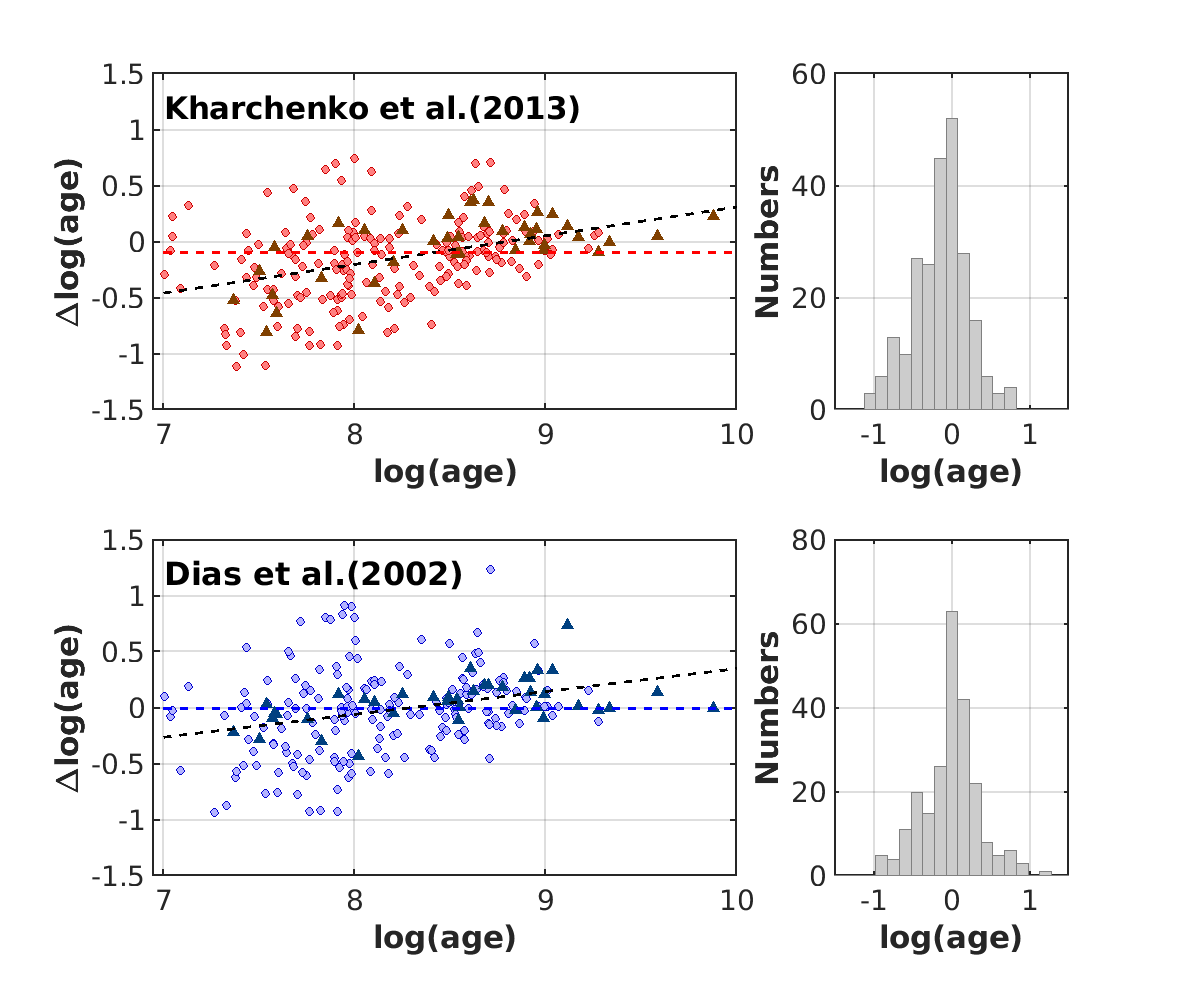}} 
    \caption{Age differences between our catalog with MWSC (upper panel) and DAML (lower panel). Triangles are clusters with spectroscopic determination of \feh. Dashed lines correspond to the  median deviations. Right panel histograms (in grey) show the difference in age distribution between our sample and the two catalogs. } 
    \label{fig:comp_lett}  
\end{figure*}

\begin{figure*}
    \centering
    \resizebox{\hsize}{!}{\includegraphics{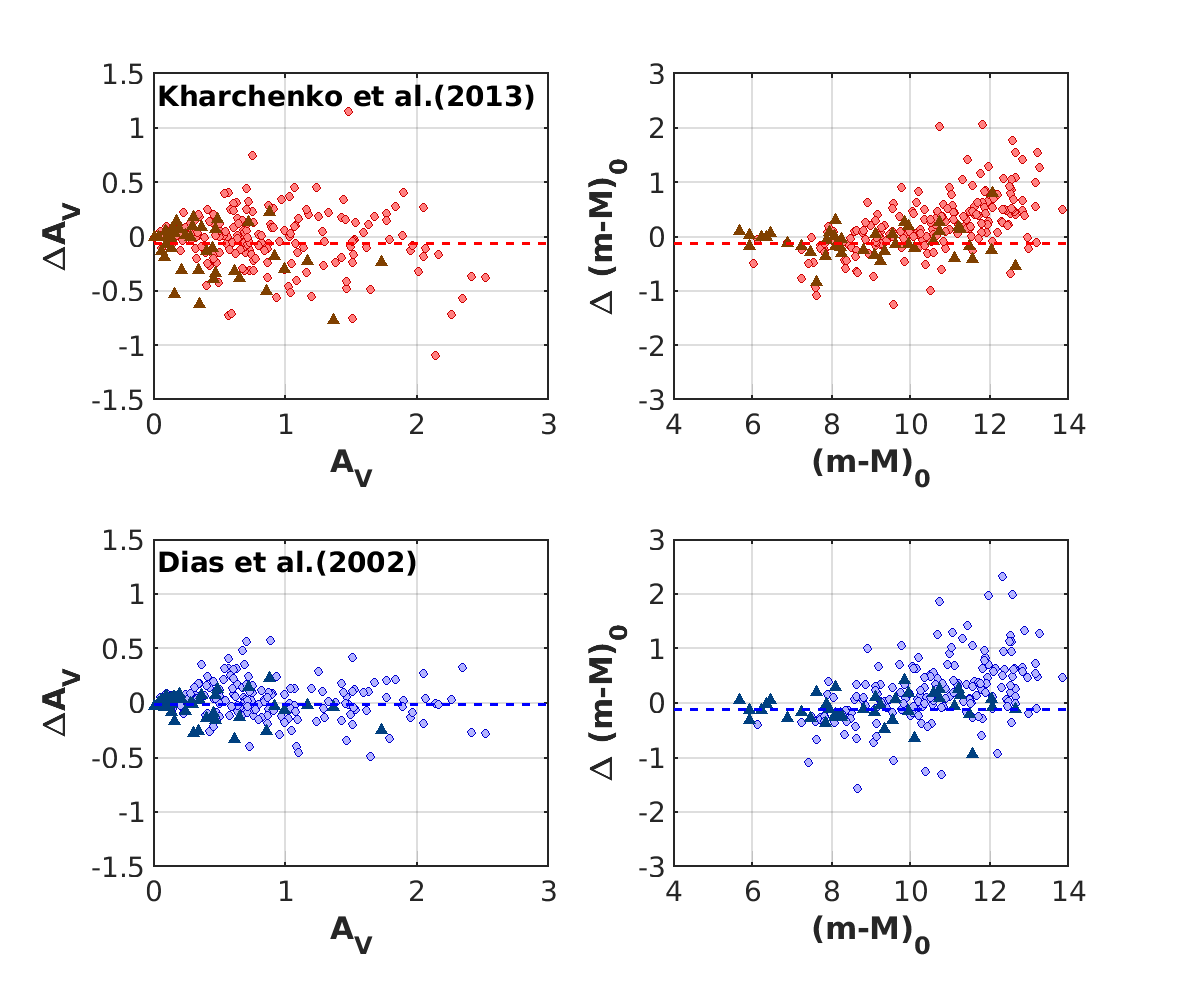}} 
    \caption{Extinction and distance modulus differences between our catalog and MWSC (upper panel) and DAML (lower panel). Triangles are clusters with spectroscopic determination of \feh. Dashed lines correspond to the  median deviations from each cluster.} 
    \label{fig:comp_lett_AV_D}  
\end{figure*}

We have 242 clusters in common with MWSC and 234 with DAML.
Fig.~\ref{fig:comp_lett}, Fig.~\ref{fig:comp_lett_AV_D}  and Table~\ref{tab:agedev} show the difference among our determination of age, extinction, and distance for those clusters included in these catalogs.

The median value of the age distribution is mainly consistent with DAML catalog, and shows a systematic of $\sim -0.09$ with MWSC. The dispersion is very large in both cases, especially for young clusters, but it is smaller for objects belonging to the HRS group. 
In addition a clear trend of the age difference with age is present, in the sense that results from \base\ are generally younger for OCs below $\log(age)<8.5$.   
These deviations are not surprising and might be ascribed to the quality of the cluster membership determination. 
Previous membership determinations are based on ground based photometry and/or proper motions and are severely hampered by field star contamination.
This problem is particularly age-related. In fact, while old clusters can count on better populated features (MS, RGB, and RC) that help the age determination from isochrone fitting, in young clusters the fit is generally based on the luminosity of the MSTO, which may be not well defined, due to the lack in the number of bright near-TO stars. In such a scenario, a different determination of membership, with the addition of bright TO stars, can change the estimation of the age (as we already saw in Sect.~\ref{sec:CC} for NGC~6793).

Fig.~\ref{fig:comp_lett_AV_D} compares globally our estimates of $A_V$ and $(m-M)_0$ with the MWSC and DAML.
Globally, no systematic, or a very small one, is present between this work and the literature concerning the value of $A_V$, but with a large dispersion. 
The distance modulus exhibits a median difference $(m-M)_{0,this work}-(m-M)_{0,lit} \sim -0.1$ for both catalogs, getting worse at $(m-M)_{0}>10$, where it becomes  $\sim 0.37, 0.27$ for MWSC and DAML catalogs respectively.

\begin{figure*}
    \centering
    \resizebox{\hsize}{!}{\includegraphics{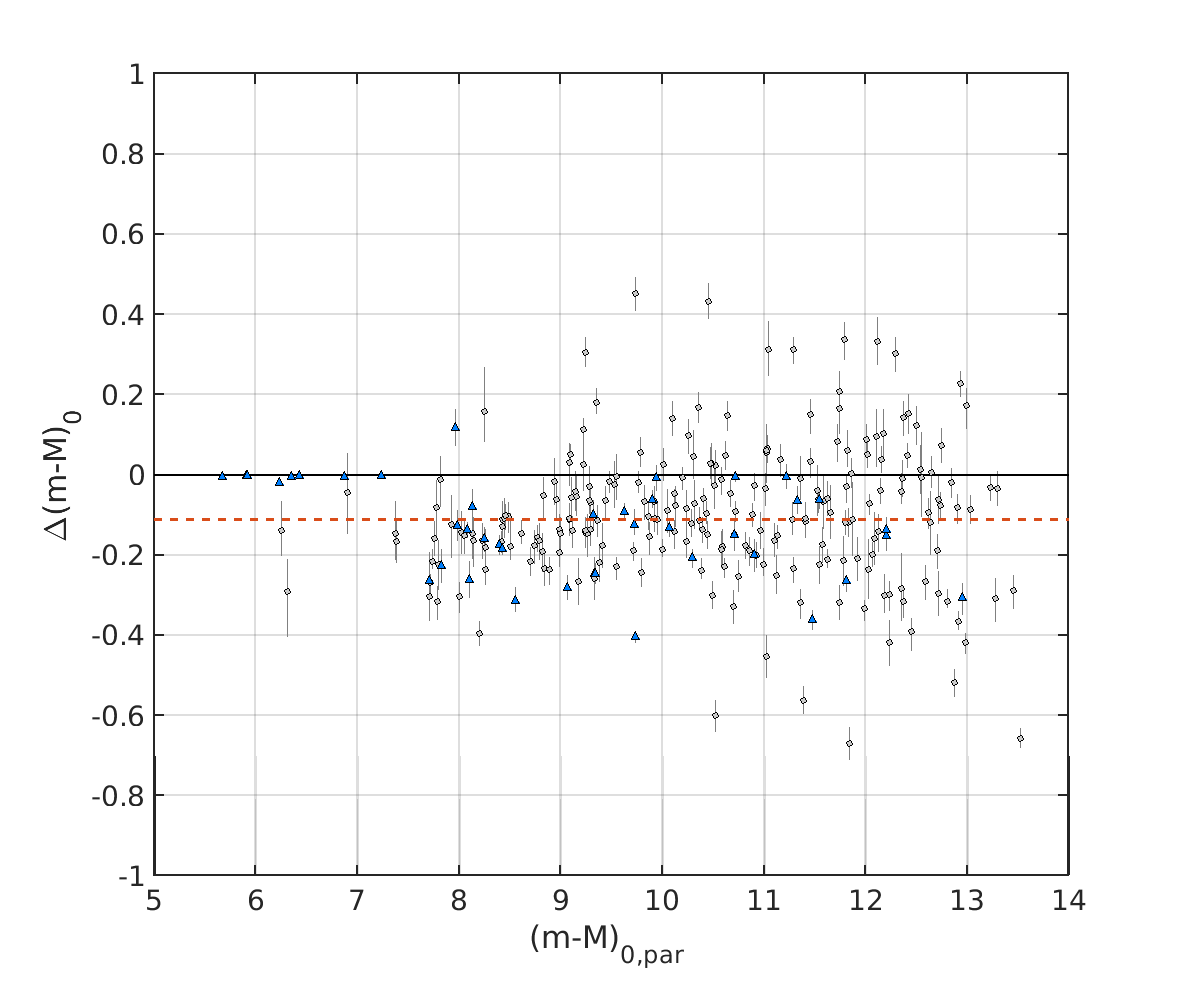}} 
        \caption{Differences of the distance modulus between posteriors ($(m-M)_\mathrm{0,med}$) and priors ($(m-M)_\mathrm{0,par}$), where the first have been derived from \base\ output and the latter from Eq.\ref{eq:parallax}. Blue triangles are clusters with spectroscopic determination of \feh, while the red dashed line is the median deviation corresponding to $\Delta(m-M)_\mathrm{0,med-par}=-0.11$ mag.}\label{fig:dist_med-par}    
\end{figure*}

{Fig.~\ref{fig:dist_med-par} shows the difference between the distance moduli derived from the analysis of \base\ results ($(m-M)_\mathrm{0,med}$, i.e. the posteriors) and from the inversion of the median parallax ($(m-M)_\mathrm{0,par}$, see Eq.\ref{eq:parallax}).
 We derive a median offset of ($(m-M)_\mathrm{0,\base\ }-(m-M)_\mathrm{0,par}=-0.11$.
 As already discussed in previous sections, the inverse of the parallax tends to overestimate the distance modulus. This is specially true when the relative uncertainty on the parallax is higher than 20\%, but it holds also when the uncertainties are lower than that. Here the majority of the clusters are more distant than 1 Kpc, in a regime where the uncertainties on the single star parallaxes are higher than 20\%. Averaging the uncertainties on the number of stars in a cluster does not reduce systematics and correlations.
The offset we find  corresponds to a medium offset of +0.021~mas in parallax.
This value is in good agreement with the well-known systematic found in \gaia\ parallaxes and reported in  \citet{Lindegren18}.

In addition, the  results show a large dispersion, with differences up to $\pm 0.5$ mag.
We cannot exclude that this due to some other effects, such as uncertainties on the extinction coefficients, or  on the assumptions on the metal content.  Stochastic effects on the Color-Magnitude diagrams of the less populated clusters can also play a significant role, as well as effects related to stellar evolution (rotation, convection)
and binarity.}

\begin{table*}
  \caption{Comparison of the age, $A_V$, and distance modulus determination in the present work with literature catalogs, namely DAML and MWSC.
  The differences are always parameters in this work - the corresponding quantity in the literature. For each difference, we list the median value and the MAD. \label{tab:agedev}}
  \tiny{\begin{center}
    \begin{tabular}{@{}lcccccc}               
    \hline  Cat.  &    $\Delta(\log(age)$    &   MAD$_{\Delta\log(age)}$ & $\Delta A_V $ &MAD$_{\Delta A_V }$ & $\Delta (m-M)_{0}$ &MAD$_{\Delta (m-M)_{0}}$ \\
\hline  
  DAML all &  0.00  & 0.17  &-0.01  & 0.07  & -0.11  &  0.29\\
  MWSC all & -0.09  & 0.19  &-0.05  & 0.13  & -0.08  &  0.28\\
  DAML HRS &  0.05  & 0.11  &-0.01  & 0.06  & -0.11  &  0.15 \\
  MWSC HRS &  0.04  & 0.10  &-0.06  & 0.14  & -0.11  &  0.15 \\  \hline
  \end{tabular}
  \end{center}}
\end{table*}

\section{Conclusions}\label{sec:conclusions}

In this work we make use of an automated method based on Bayesian classification, \base, to derive the age of 269 OCs using GRD2 photometry.
The parameter determination precision is
$\sigma_{\log{(age)}}=0.100$, $\sigma_{A_V}=0.033$, and $\sigma_{(m-M)_0}=0.037$, while, their medians are $\widetilde{\sigma}_{\log{(age)}}=0.024$, $\widetilde{\sigma}_{A_V}=0.023$, and $\widetilde{\sigma}_{(m-M)_0}=0.025$.
In all the calculations we assume a fixed value of the metallicity \feh, taking it either from high or low resolution spectroscopy or from photometry. When no information is available, we assume $\feh=0.0$.
We discuss the effect that the prior  has on our results through a series of numerical sensitivity experiments. We find that in the worst case (no information on \feh), we have a
$\Delta\log(age)= \pm 0.05$.
Comparing our results with existing literature data, we find a large dispersion on age, and $A_V$ with no or a little systematics. On average younger ages are affected by large differences with existing catalogs. This could well be due to the high quality of the \gaia\ data, i.e. more reliable membership determination and photometry. 
However, we cannot exclude that \base\  tends to underestimate the ages of young clusters.  
We point out that this is the largest data base of OC parameters derived using homogeneous and high quality data and this  method. 
In this work we make use only of the information from the  three \gaia\ bands. This is motivated by the high quality of the \gaia\ photometry. 
However, \base\  runs show that using only these magnitudes is not possible to resolve the degeneracy between the four cluster parameters, mainly the distance modulus and the extinction {\citep[see also][]{Andrae18}}. 
For this reason we have analyzed only low extinction objects. A further development will be to use information from complementary photometry to alleviate the degeneracy and extend the present catalog to higher extinction regimes.

\section*{Acknowledgements}
This work makes use of data products from: the ESA \gaia\ mission (\url{gea.esac.esa.int/archive/}), funded by national institutions participating in the \gaia\ Multilateral Agreement.
This work was supported by ASI (Italian Space Agency) under contract 2014-025-R.1.2015.
AB acknowledges funding from PREMIALE 2015 MITiC.
This work was supported by the MINECO (Spanish Ministry of Economy) through grant ESP2016-80079-C2-1-R (MINECO/FEDER, UE) and MDM-2014-0369 of ICCUB (Unidad de Excelencia 'María de Maeztu'). 
C.S. and L.C. acknowledge support from the CNES and from the "programme national cosmologie et galaxies" (PNCG) of CNRS/INSU.
The Portuguese Fundação para a Ciência e a Tecnologia (FCT) through the Strategic Programme UID/FIS/00099/2013 for CENTRA"
UH acknowledges support from the Swedish National Space Agency (SNSA/Rymdstyrelsen).

\bibliographystyle{aa} 
\linespread{1.5}		
\bibliography{biblio}

\begin{appendix}
\section{CLUSTER TABLES}
\def\arraystretch{0.9}
\begin{table*}[!htbp]
\caption{\label{Table_spectro} {List of the first 10 of the 44 HRS+LRS clusters (the complete list is available electronic version of this paper). The values of $\log(age)$, distance moduli, and extinctions $A_V$ correspond to the median of each \base\ posterior distribution. $A_{G_{TO}}$ column reports the extinction in \Gmag\ at the turnoff of each cluster. Uncertainties are given and correspond to the 16$^\mathrm{th}$ (superscript) and 84$^\mathrm{th}$ (subscript) percentiles of posterior distribution. \feh column lists the metallicity used in \base\ computations.}}
\tiny{\begin{center}
\begin{tabular}{@{}l r r r r r r r}           
\hline
\input{R_FINAL_TABLE_FIX_HRS+LRS_SHORT10.tex}
\end{tabular}
\end{center}}
\end{table*}

\def\arraystretch{0.9}
\begin{table*}
\caption{\label{Table_phot} {List of the first 10 of the 46 PHS clusters (the complete list is available electronic version of this paper). Columns are the same as in Table~\ref{Table_spectro}. }}
\tiny{\begin{center}
\begin{tabular}{@{}l r r r r r r r}           
\hline
\input{R_FINAL_TABLE_FIX_PHS_SHORT10.tex}
\end{tabular}
\end{center}}
\end{table*}

\def\arraystretch{0.9}
\begin{table*}
\caption{\label{Table_noinfo} {List of the first 10 of the 179 NC clusters (the complete list is available electronic version of this paper). Columns are the same as in Table~\ref{Table_spectro}.}}
\tiny{\begin{center}
\begin{tabular}{@{}l r r r r r r r}           
\hline
\input{R_FINAL_TABLE_FIX_NC_SHORT10.tex}
\end{tabular}
\end{center}}
\end{table*}
\end{appendix}
\end{document}

%% file: newcommand.tex
    \newcommand{\D}{\mathrm{d}} 

    \newcommand{\feh}{\mbox{[Fe/H]}}		
    \newcommand{\dnu}{\mbox{$\Delta\nu$}}		

    \newcommand{\Dnu}{\dnu}		
    \newcommand{\numax}{\mbox{$\nu_\mathrm{max}$}}   

    \newcommand{\Msun}{\mbox{M$_{\odot}$}}		                

    \newcommand{\BR}{\mbox{$(G_{\rm BP}\!-\!G_{\rm BP})$}}	

    \newcommand{\Gmag}{\mbox{$G$}}
    \newcommand{\Bp}{\mbox{$G_{\rm BP}$}}
    \newcommand{\Rp}{\mbox{$G_{\rm RP}$}}

    \newcommand{\AG}{\mbox{$A_G$}}     
    \newcommand{\ABp}{\mbox{$A_{G_{\rm BP}}$}}
    \newcommand{\ARp}{\mbox{$A_{G_{\rm RP}}$}}

    \newcommand{\beq}{\begin{equation}}
    \newcommand{\eeq}{\end{equation}}	
    \newcommand{\beqa}{\begin{eqnarray}}
    \newcommand{\eeqa}{\end{eqnarray}}	
    \newcommand{\bitem}{\begin{itemize}}
    \newcommand{\eitem}{\end{itemize}}	
    \newcommand{\benum}{\begin{enumerate}}	
    \newcommand{\eenum}{\end{enumerate}}	
    

%% file: R_FINAL_TABLE_FIX_HRS+LRS_SHORT10.tex
cluster              & ra         & dec        & $\log (age)$ & $(m-M)_0$  & $A_V$      & $A_{G_{TO}}$ & \feh      \\ \hline
Blanco1              &   +0.853 &    0.853 &  7.98$^{8.00}_{7.94}$ &  6.88$^{6.88}_{6.88}$ &  0.03$^{0.03}_{0.03}$ &  0.03$^{0.03}_{0.03}$ &  0.000\\
IC2391               & +130.292 &  130.292 &  7.56$^{7.58}_{7.54}$ &  5.91$^{5.91}_{5.91}$ &  0.09$^{0.09}_{0.09}$ &  0.09$^{0.09}_{0.09}$ &  0.000\\
IC2602               & +160.613 &  160.613 &  7.55$^{7.56}_{7.53}$ &  5.91$^{5.91}_{5.91}$ &  0.10$^{0.10}_{0.10}$ &  0.09$^{0.09}_{0.09}$ &  0.000\\
IC2714               & +169.373 &  169.373 &  8.55$^{8.55}_{8.55}$ &  10.71$^{10.72}_{10.70}$ &  0.99$^{1.00}_{0.98}$ &  0.97$^{0.98}_{0.96}$ &  0.020\\
IC4665               & +266.554 &  266.554 &  7.58$^{7.64}_{7.55}$ &  7.45$^{7.49}_{7.39}$ &  0.40$^{0.43}_{0.36}$ &  0.39$^{0.42}_{0.35}$ &  -0.030\\
IC4756               & +279.649 &  279.649 &  8.99$^{8.99}_{8.97}$ &  8.40$^{8.40}_{8.40}$ &  0.40$^{0.40}_{0.40}$ &  0.39$^{0.39}_{0.39}$ &  0.000\\
Melotte20            &  +51.617 &   51.617 &  7.75$^{7.76}_{7.74}$ &  6.21$^{6.21}_{6.21}$ &  0.28$^{0.28}_{0.28}$ &  0.27$^{0.27}_{0.27}$ &  0.140\\
Melotte22            &  +56.601 &   56.601 &  7.94$^{7.97}_{7.92}$ &  5.67$^{5.67}_{5.67}$ &  0.14$^{0.14}_{0.14}$ &  0.14$^{0.14}_{0.14}$ &  0.000\\
Melotte71            & +114.383 &  114.383 &  9.11$^{9.11}_{9.11}$ &  11.55$^{11.56}_{11.54}$ &  0.48$^{0.49}_{0.47}$ &  0.47$^{0.48}_{0.46}$ &  -0.270\\
NGC0188              &  +11.798 &   11.798 &  9.69$^{9.69}_{9.68}$ &  11.49$^{11.49}_{11.49}$ &  0.26$^{0.26}_{0.26}$ &  0.26$^{0.26}_{0.26}$ &  0.000\\

%% file: R_FINAL_TABLE_FIX_PHS_SHORT10.tex
cluster              & ra         & dec        & $\log (age)$ & $(m-M)_0$  & $A_V$      & $A_{G_{TO}}$ & \feh      \\ \hline
Alessi5              & +160.819 &  160.819 &  7.72$^{7.74}_{7.61}$ &  7.72$^{7.78}_{7.69}$ &  0.59$^{0.62}_{0.56}$ &  0.58$^{0.60}_{0.55}$ &  -0.382\\
Alessi6              & +220.058 &  220.058 &  8.78$^{8.80}_{8.74}$ &  9.55$^{9.58}_{9.53}$ &  0.72$^{0.74}_{0.70}$ &  0.70$^{0.72}_{0.68}$ &  -0.154\\
Alessi24             & +260.764 &  260.764 &  7.95$^{7.97}_{7.91}$ &  8.30$^{8.33}_{8.28}$ &  0.34$^{0.36}_{0.31}$ &  0.33$^{0.35}_{0.30}$ &  -0.133\\
BH99                 & +159.553 &  159.553 &  7.91$^{7.94}_{7.88}$ &  8.08$^{8.11}_{8.06}$ &  0.20$^{0.23}_{0.18}$ &  0.20$^{0.22}_{0.17}$ &  0.000\\
Collinder140         & +110.882 &  110.882 &  7.47$^{7.52}_{7.43}$ &  7.81$^{7.86}_{7.76}$ &  0.10$^{0.13}_{0.08}$ &  0.10$^{0.12}_{0.08}$ &  0.010\\
Czernik27            & +105.830 &  105.830 &  9.06$^{9.08}_{9.04}$ &  12.97$^{13.00}_{12.93}$ &  0.54$^{0.57}_{0.51}$ &  0.52$^{0.56}_{0.49}$ &  -0.380\\
Harvard5             & +186.817 &  186.817 &  7.81$^{8.00}_{7.75}$ &  10.40$^{10.43}_{10.38}$ &  0.68$^{0.71}_{0.65}$ &  0.66$^{0.69}_{0.64}$ &  -0.090\\
IC1369               & +318.033 &  318.033 &  8.46$^{8.46}_{8.46}$ &  12.54$^{12.56}_{12.53}$ &  2.05$^{2.06}_{2.04}$ &  2.00$^{2.01}_{1.99}$ &  0.090\\
IC2488               & +141.857 &  141.857 &  8.20$^{8.22}_{8.15}$ &  10.63$^{10.64}_{10.61}$ &  0.70$^{0.72}_{0.69}$ &  0.69$^{0.70}_{0.67}$ &  0.080\\
IC4725               & +277.937 &  277.937 &  7.98$^{7.99}_{7.97}$ &  9.16$^{9.18}_{9.14}$ &  1.09$^{1.10}_{1.07}$ &  1.06$^{1.08}_{1.05}$ &  0.000\\

%% file: R_FINAL_TABLE_FIX_NC_SHORT10.tex
cluster              & ra         & dec        & $\log (age)$ & $(m-M)_0$  & $A_V$      & $A_{G_{TO}}$ & \feh      \\ \hline
ASCC6                &  +26.846 &   26.846 &  7.68$^{7.69}_{7.66}$ &  11.10$^{11.12}_{11.08}$ &  0.87$^{0.89}_{0.85}$ &  0.85$^{0.87}_{0.83}$ &  0.000\\
ASCC10               &  +51.870 &   51.870 &  8.60$^{8.64}_{8.50}$ &  8.91$^{8.95}_{8.88}$ &  0.44$^{0.48}_{0.41}$ &  0.43$^{0.46}_{0.40}$ &  0.000\\
ASCC13               &  +78.255 &   78.255 &  7.65$^{7.66}_{7.63}$ &  10.15$^{10.18}_{10.12}$ &  0.68$^{0.70}_{0.65}$ &  0.66$^{0.68}_{0.64}$ &  0.000\\
ASCC16               &  +81.198 &   81.198 &  7.05$^{7.05}_{7.05}$ &  7.52$^{7.54}_{7.51}$ &  0.10$^{0.11}_{0.08}$ &  0.09$^{0.11}_{0.08}$ &  0.000\\
ASCC19               &  +81.982 &   81.982 &  7.09$^{7.09}_{7.09}$ &  7.47$^{7.50}_{7.45}$ &  0.06$^{0.07}_{0.04}$ &  0.06$^{0.06}_{0.04}$ &  0.000\\
ASCC21               &  +82.179 &   82.179 &  7.04$^{7.04}_{7.03}$ &  7.41$^{7.44}_{7.39}$ &  0.12$^{0.14}_{0.10}$ &  0.12$^{0.14}_{0.10}$ &  0.000\\
ASCC22               &  +93.656 &   93.656 &  8.55$^{8.60}_{8.49}$ &  9.54$^{9.58}_{9.51}$ &  0.55$^{0.59}_{0.52}$ &  0.54$^{0.57}_{0.51}$ &  0.000\\
ASCC23               &  +95.047 &   95.047 &  8.48$^{8.50}_{8.44}$ &  8.85$^{8.88}_{8.83}$ &  0.28$^{0.31}_{0.26}$ &  0.28$^{0.30}_{0.26}$ &  0.000\\
ASCC29               & +103.571 &  103.571 &  7.95$^{7.98}_{7.91}$ &  10.05$^{10.08}_{10.02}$ &  0.24$^{0.27}_{0.21}$ &  0.23$^{0.26}_{0.21}$ &  0.000\\
ASCC32               & +105.714 &  105.714 &  7.40$^{7.40}_{7.40}$ &  9.32$^{9.34}_{9.29}$ &  0.22$^{0.23}_{0.20}$ &  0.21$^{0.23}_{0.19}$ &  0.000\\